\documentclass[aps,prb,superscriptaddress,twocolumn]{revtex4-1}
\usepackage{lmodern}

\usepackage[latin9]{inputenc}
\usepackage{amsmath}
\usepackage{amssymb}
\usepackage{graphicx}
\usepackage{epstopdf}
\usepackage{color}
\usepackage{enumerate}
\usepackage{ulem}

\begin{document}

\title{The discovery, disappearance and re-emergence of radiation-stimulated superconductivity}
\author{T. M.  Klapwijk}
\email{t.m.klapwijk@tudelft.nl}

\affiliation{Kavli Institute of Nanoscience, Delft University of Technology, Delft, The Netherlands}
\affiliation{Institute for Topological Materials, Julius Maximilian University of Würzburg, Würzburg, Germany}
\affiliation{Physics Department, Moscow State University of Education, Moscow, Russia }
\author{P.J. de Visser}
\affiliation{SRON Netherlands Institute for Space Research, Sorbonnelaan 2, 3584CA Utrecht, The Netherlands}
\date{\today}

\begin{abstract}
We trace the historical fate of experiment and theory of microwave-stimulated superconductivity as originally reported for constriction-type superconducting weak links. It is shown that the observed effect disappeared by improving weak links to obtain the desired Josephson-properties. Separate experiments were carried out to evaluate the validity of the proposed theory of Eliash'berg for \textit{energy-gap-enhancement} in superconducting films in a microwave field, without reaching a full quantitatively reliable measurement of the stimulated energy gap in a microwave-field, but convincing enough to understand the earlier deviations from the Josephson-effect. Over the same time-period microwave-stimulated superconductivity continued to be present in superconductor-normal metal-superconductor Josephson weak links. This experimental body of work was left unexplained for several decades and could only be understood properly after the microscopic theory of the proximity-effect had matured enough, including its non-equilibrium aspects. It implies that the increase in critical current in weak-link Josephson-junctions is due to an \textit{enhancement of the phase-coherence} rather than to an \textit{enhancement of the energy-gap} as proposed by Eliash'berg. The complex interplay between proximity-effect and the occupation of states continuous to be, in a variety of ways, at the core of the ongoing research on hybrid Josephson-junctions. The subject of radiation-enhanced superconductivity has re-emerged in the study of the power-dependence of superconducting microwave resonators, but also in the light-induced emergence of superconductivity in complex materials.
\end{abstract}
\maketitle

\vspace{0.5cm}
\section{Introduction: discovery}\label{One}
\begin{figure}[t!]
   	\includegraphics[width=1\linewidth, angle=90] {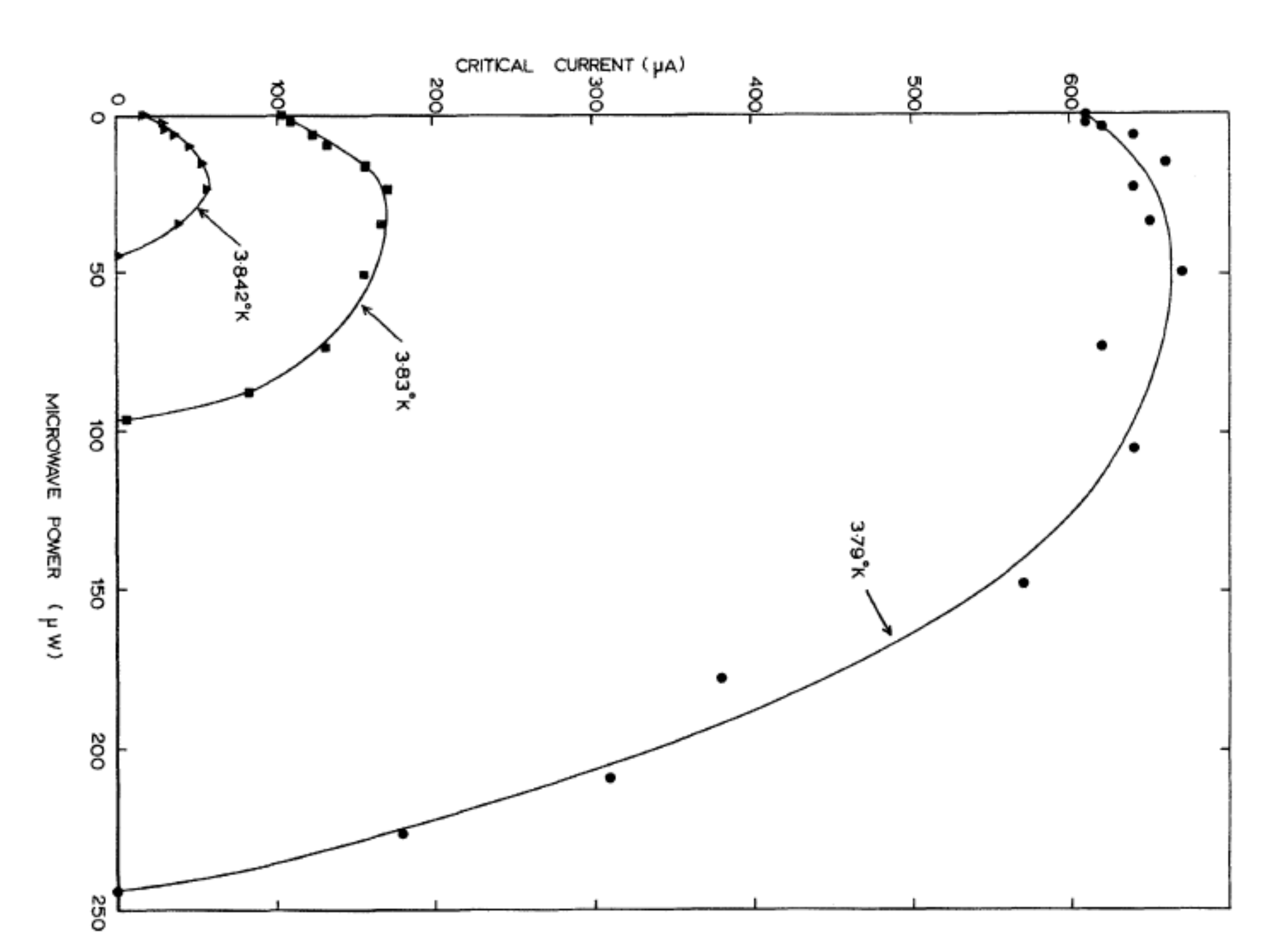}   		
	\caption{First published observation of a microwave-enhanced critical supercurrent of a microbridge (in 1966). (Taken from Wyatt et al\cite{Wyatt1966})}
   	\label{Fig1}
   \end{figure}

Anderson\cite{AndersonDayem1964} had early on the wonderful insight that the Josephson-effect was so universal that any type of weak link between two superconductors, including a constriction in a superconducting film would show Shapiro-steps, as observed in the current-voltage characteristics of superconducting tunnel-junctions\cite{Shapiro1963}. This basic phenomenon, current-steps at quantized values of the voltage, was  quickly proven to be present in the experiment published by Anderson and Dayem\cite{AndersonDayem1964}. One of the beauties is that these ideas can be applied to other quantum fluids such a $^4He$ and $^3He$. An early attempt on $^4He$, was carried out by Richards and Anderson\cite{Richards1965}, although it took much longer to turn this type of experiments into a convincing result\cite{Varoquaux2015}, using superfluid $^3He$. But even for the superconducting weak link some confusion arose, as became clear in a subsequent publication\cite{Dayem1967} about the earlier experiment:
\begin{quote}
\small{
\textit
{We also observed a baffling effect, namely, the increase of the critical supercurrent of the bridge with the applied microwave field for frequencies larger than 2 GHz.}} 
\end{quote}
In other words, the amplitude of the Josephson-current, at $V=0$, which is supposed to decrease according to the 0-th order Bessel-function\cite{Shapiro1964} \textit{increases} with microwave power.  These results were probably observed in the early stage of the work on microbridges, but published by Dayem and Wiegand\cite{Dayem1967} much later in 1967, after an earlier published result of Wyatt et al\cite{Wyatt1966}, shown in Fig.\ref{Fig1}.

Research on superconducting microbridges developed in parallel with superconducting pointcontacts, usually made of niobium, and more suitable for high-frequency experiments because of a higher impedance, better matched to free space, and analogous to whisker-diodes used in radio-receiver techniques\cite{GrimesPRB1968}.   
The research on microbridges was taken over in the early 70-ies by Gregers-Hansen and Levinsen  at the Ørsted Institut in Copenhagen (Fig.\ref{Fig2}). Simultaneously, work on microbridges was carried out at the University of Kharkov in the Ukraine, led by V. M. Dmitriev\cite{Dmitriev1971}, who was exposed to this topic when working with Wyatt\cite{Wyatt1966} at the University of Nottingham. In this work the devices were reported to be on short tin (Sn) bridges of less than 3 to 4 micrometer wide. 

In parallel to these experimental developments, the microscopic theory of superconductivity was being developed at the Landau-institute in Chernogolovka in Russia. A striking prediction was articulated by Eliash'berg in 1969\cite{Eliashberg1970}. In the title it says \textit{Film superconductivity stimulated by a high-frequency field}. It is a theory for the superconducting energy gap $\Delta$, and points out that solutions exists for $T>T_c$, but whether they are stable or metastable is left open. It also states that the gap-enhancement will also lead to an increase of the critical current, with the sentence: \textit{In all probability, this is precisely the phenomenon observed by Dayem and Wiegand.} In a subsequent article by Ivlev\cite{Ivlev1971}, the kinetic equation is given together with a calculation of the critical pair-breaking current. At that point in time,  also a reference to Wyatt et al\cite{Wyatt1966} is included and a reference to an announced experimental contribution from Dmitriev at al\cite{Dmitriev1971}, including the claim that superconductivity is found above $T>T_c$. In a subsequent letter Ivlev and Eliashberg\cite{Ivlev1971b} present the kinetic equation, as well as the historically correct sequence of references of Wyatt et al\cite{Wyatt1966}, followed by Dayem and Wiegand\cite{Dayem1967}, and then the new article by Dmitriev et al\cite{Dmitriev1971}. The full theoretical development was later made more accessible to the international scientific community through a publication by Ivlev, Lisitsyn, and Eliash'berg\cite{Ivlev1973} in the Journal of Low Temperature Physics.           

The conclusion is that Eliash'berg and co-workers gradually interpreted their theoretical work as being applicable to the observations reported for superconducting microbridges, although these structures were at the same time increasingly viewed as exhibiting the Josephson-effect, certainly in the light of the theory put forward by Aslamazov and Larkin\cite{AL1968} in 1968. Moreover, the research on superconducting pointcontacts were with respect to the microwave response in agreement with the current-biased RSJ-model\cite{Russer1970,RusserJAP1972} and did not show any trace of the Dayem-Wyatt effect\cite{GrimesPRB1968}. So it appeared that the Dayem-Wyatt effect was in some way typical for superconducting microbridges.    

\section{A nano-route to improve the Josephson-effect}\label{Two}

\begin{figure}[t!]
   	\includegraphics[width=1\linewidth] {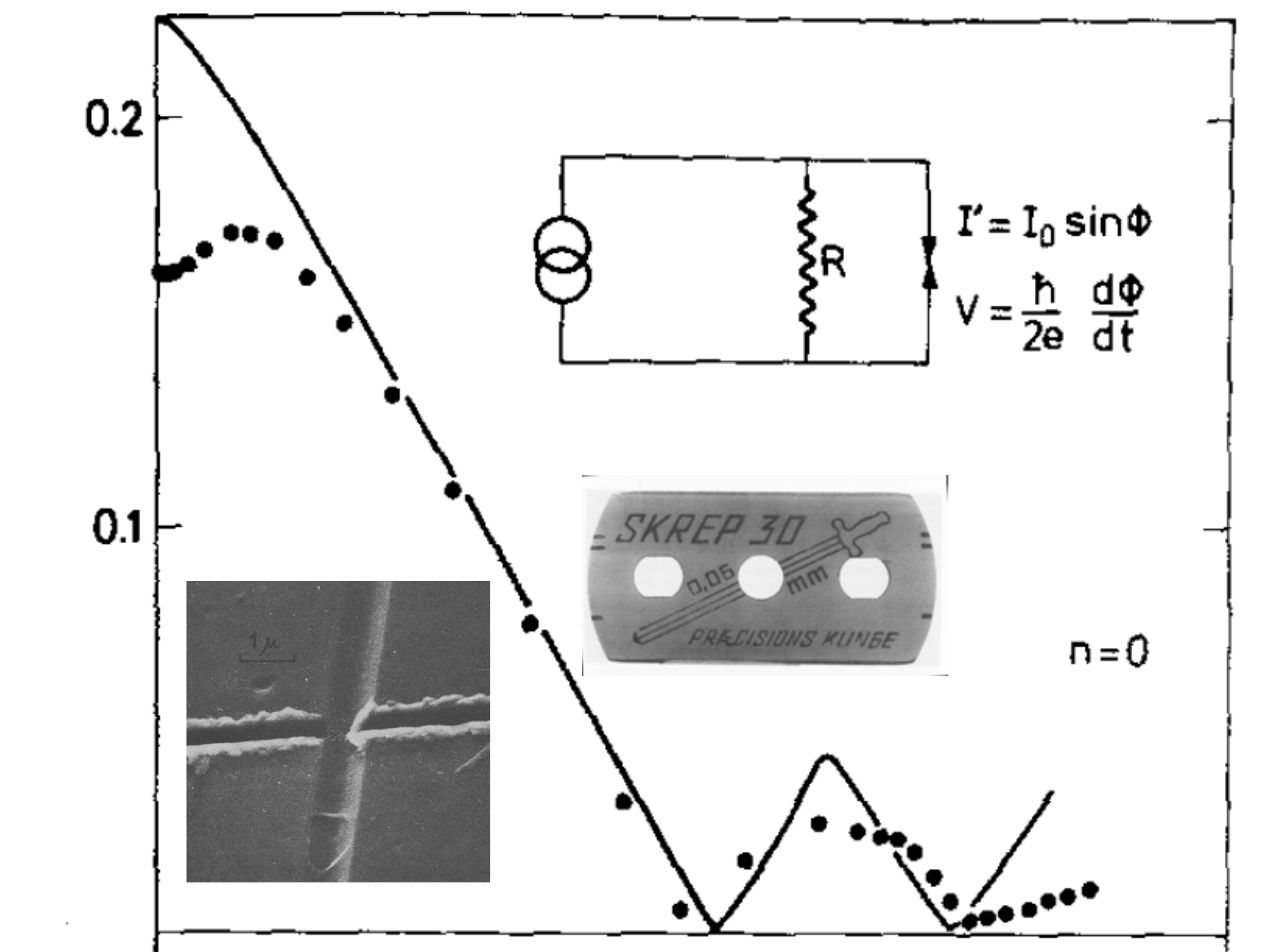}   		
	\caption{Critical supercurrent as a function of microwavepower for a razorblade-fabricated nanoscale microbridge of tin (Sn). The circuit in the inset represents the resistively-shunted junction model with a current source, acting as a dc source and an ac source to model the microwaves. The full line are the predictions from this RSJ-model. Note the presence of a relatively small enhancement of the critical current. (Taken from Gregers-Hansen and Levinsen\cite{GregersHansenSSC1971,GregersHansenLevinsenPRL1971})}
	\label{Fig2}
   \end{figure}

The theoretical understanding of the Josephson-effect in constriction-type microbridges was initiated by Aslamazov and Larkin\cite{AL1968}. They recognized that if the dimensions of a superconducting contact would be shorter and narrower than the coherence length $\xi$, the problem could be reduced  to a solution from the microscopically justified Ginzburg-Landau equations for a 'dirty' superconductor. The dominant term in the free energy would become the gradient of the order parameter leading to a search for solutions from the Laplace equation for the order parameter $\Delta$: 
\begin{equation}
\nabla^2\Delta=0
\label{AL}
\end{equation}    
Similarly, the normal state resistance follows for diffusive transport from the Laplace equation for the electrostatic potential, which led them to propose the simple circuit-diagram, shown in Fig.\ref{Fig2}, known as the resistively-shunted junction (RSJ-)model. It consists of a normal resistance in parallel with a circuit element which fulfils the two Josephson-equations, $I_s=I_c\sin\phi$ and $d\phi/{dt}=2eV/\hbar$. The same model was independently postulated by Stewart\cite{Stewart1968} and McCumber\cite{McCumber1968} as an engineering approach to understand a low capacitance point-contact-like Josephson-junction, with its much less prominent presence of hysteresis. Some theoretical results are shown in Fig. \ref{Fig2}, taken from Gregers-Hansen and Levinsen\cite{GregersHansenSSC1971,GregersHansenLevinsenPRL1971},  of the expected response of the critical current to microwave-radiation, which is assumed to be applied as a current-source with a dc and an ac component. The vertical axis is in $mA$ and the horizontal axis, the microwave amplitude in arbitrary units. For the critical current one expects an oscillatory dependence on microwave power, roughly like the zeroth order Bessel function. 

In the early 70-ies the experimental results on superconducting microbridges were emerging from the group at the Ørsted Institute in Copenhagen\cite{GregersHansenSSC1971,GregersHansenLevinsenPRL1971}. Fig. \ref{Fig2} shows a measurement of the critical current of a tin (Sn) microbridge as a function of microwave power.  It shows the oscillatory pattern as expected based on the Bessel-functions. Secondly, in contrast to previous work, which served as an experimental frame of reference for Eliash'berg and co-workers the group in Copenhagen developed a new technique to make constriction-type microbridges, which were shorter and narrower than the previously used microbridges. Gregers-Hansen and Levinsen\cite{GregersHansenLevinsenPRL1971} describe their method in a footnote as follows:    

\begin{quotation}
\textit{After a light cut with a razor blade in the surface of a glass substrate, the substrate was immersed in dilute hydrofluoric acid for a short time. In this way a very regular groove about $0.5~\mu m$ wide and of approximately semicircular cross section was obtained. After evaporation of a $0.1~\mu m$-thick tin layer another cut with a razor blade was made crossing the groove. The razor blade removed the tin film along a line often only $0.1~\mu m$ wide leaving the tin in the bottom of the groove as the bridge connecting the two halves of divided film.}
\end{quotation}
In Fig.~\ref{Fig2} the inset shows a SEM-picture of a completed device as well as the type of razorblade used. The authors observed that in contrast to the earlier results on microbridges  the behavior was more in agreement with the Josephson-effect and the microwave-enhancement had become much less dominant. In order to understand the results quantitatively analog computer calculations of the current-biased resistively-shunted junction model were developed by Russer\cite{Russer1970} in Vienna. As shown in the figure, in the data there is an initial rise in the critical current, reminiscent of the Dayem-Wyatt effect, considerably reduced, but still in disagreement with the theory. Christiansen et al\cite{ChristiansenJLTP1971} addressed this disagreement  and appear to have been the first outside the Soviet Union to draw attention to Eliashberg's work\cite{Eliashberg1970}

In 1971 the claim of the Copenhagen-group was that the Josephson-behavior in microbridges, provided they were shorter and narrower than the original Anderson-Dayem bridges, were largely in agreement with the theory. Although, the coherent motion of vortices might produce similar results, the common assumption was that the constriction should have dimensions smaller than the coherence length $\xi_0$, with values of $38~nm$ for niobium, $230~nm$ for Sn, and $1.6~\mu m$ for Al. In practice for deposited thin films the coherence lengths are even shorter, $\sqrt{\xi_0\ell}$ with $\ell$ the mean free path for elastic electron scattering. These numbers provided an early push towards lithography on a nanoscale. (It is worth remembering that integrated circuits were at that time based on a scale of $5~\mu m$.) The razor-blade technology inspired Tinkham's group\cite{SkocpolJLTP1974} at Harvard to use a diamond knife, our group at Delft\cite{Mooij1974} to use a diamond-cutting tool in a dedicated scratching-apparatus, Gubankov's group at IREE in Moscow a quartz-fiber as a shadow-mask\cite{GubankovVTB1973} later replaced  by the razorblade-technique\cite{Gubankov1976}, Palmer and Decker\cite{Palmer1973} at Caltech used an inverted optical microscope, and Laibowitz and Hatzakis at IBM-Yorktown Heights an early version of an electron-beam pattern generator\cite{Laibowitz1973}. 

These experimental developments led to a range of Josephson-effect data of mixed quality, which could have several reasons. First, it could be due to the choice of materials in relation to the coherence length and the available dimensions. Secondly, it could be due to a lack of awareness that, measuring at finite voltage, as is done when studying Shapiro-steps, power is being dissipated, which addresses the choice of the substrate as well as the geometry of the device\cite{SBT1974} and thermal hysteresis\cite{Divin1974} rather than capacitive hysteresis\cite{McCumber1968}. Thirdly, the solvable theoretical problems such as provided by Aslamazov and Larkin\cite{AL1968}, and by Kulik and Omel'yanchuk\cite{KOI1975} were a reminder of the importance of the boundary conditions \textit{i.e.} the transition from equilibrium superconducting electrodes to the geometric constriction part of  the superconductor. Fourthly, the continued success of mechanical pointcontact Josephson-junctions raised questions about the difference with microbridges. This led to an evolution towards so-called variable-thickness bridges (VTB's) to improve the cooling\cite{KlapwijkVeenstra1974,KlapwijkMooij1975,Octavio1977} or to get better access to the physics\cite{GubankovVTB1977}.   

This change of device-geometry led to a considerable reduction of the earlier observed microwave-enhanced critical current and eventually the complete disappearance as a relevant topic. The current embodiment of this type of microbridges is the mechanical break junction (MBJ) \cite{Ruitenbeek1996} in which the constriction is on an atomic scale, with the equilibrium banks unavoidably geometrically more 'bulky' (see Section \ref{Five}).  In this geometry the properties are controlled by banks-in-equilibrium and the short bridge carries a supercurrent driven by the difference in the quantum phases of each of the banks. In such an arrangement the microwave field is coupled to the difference in the quantum-phases and the standard Shapiro-steps are being observed without a signature of any microwave-enhancement of the critical supercurrent. 

Apart from this constriction-type microbridges, which evolved towards the atomic scale break junctions, a separate set of microbridges was formed by the constriction-type SNS devices in which the weak link was a normal metal. These devices showed also a microwave-enhanced critical current, but embedded in a good Josephson-effect\cite{NotarysPRL1973,WarlaumontPRL1979}. These empirical observations dropped largely out of sight, but it presaged the message that there is more to it than the Eliash'berg effect to be discussed in Section \ref{Three}. We will return to the microwave-stimulated superconductivity in SNS microbridges in Section \ref{Five}. 

\begin{figure}[t!]
   	\includegraphics[width=1\linewidth, angle=0] {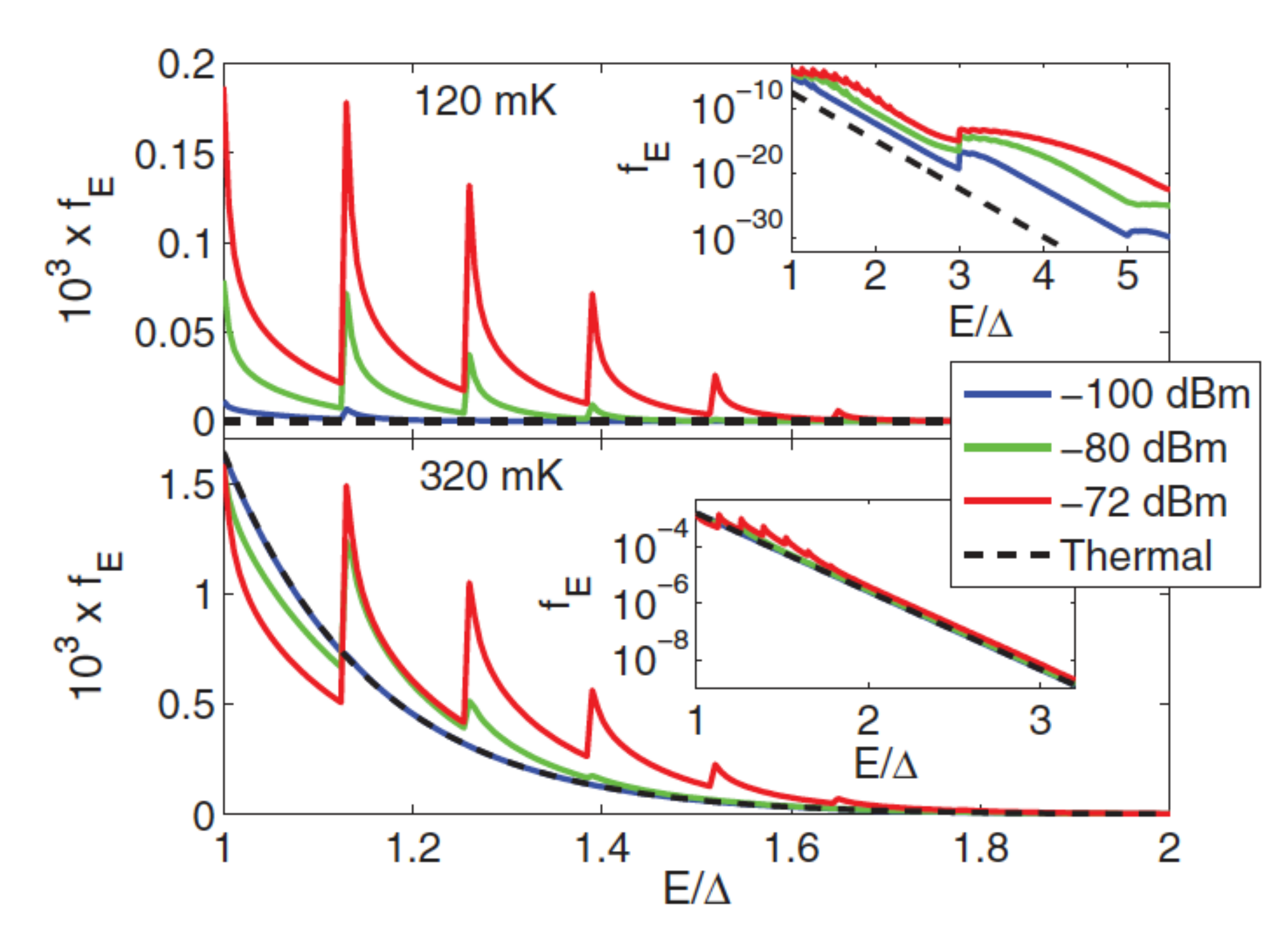}   		
	\caption{Example of microwave-induced changes of the distribution-function, $f(E)$,  of superconducting aluminium at two different temperatures, 120 mK and 320 mK, for increasing microwave power. The peaks are replicas of the density of states at $E=\Delta$ at multiples of the photon-energy $\hbar\omega$. Note the \textit{reduction} of the occupation $f_E$ between $E/\Delta=1$  and  $1+\hbar\omega/\Delta$ at 320 mK compared  to a thermal distribution, which is causing the enhancement of the energy gap. The curves in the upper panel, at 120 mK, illustrate that, at low temperatures, the Eliash'berg-effect is negligible and the dominant effect is a higher occupation at all energies (Taken from De Visser et al\cite{DeVisserPRL2014})}
   	\label{Fig3}
   \end{figure}  

\section{Eliash'berg effect}\label{Three}  
	The crucial insight of Eliash'berg\cite{Eliashberg1970} is centered around the effect of microwaves on a superconductor in view of the fact that in the BCS-theory the superconducting energy gap $\Delta$ depends on temperature by the distribution-function $f(E)$, which in equilibrium is the Fermi-Dirac distribution. With increasing temperature $T$ more and more $\vec{ k}$-states get occupied by quasiparticles, which block these states for Cooper-pair formation. The result is a fast reduction of the energy-gap with increasing $T$, eventually leading to $T_c$. This insight was expressed earlier by Parmenter\cite{Parmenter1961}, who proposed the removal of the excess quasiparticles by tunnelling, which was carried out in practice much later, in 1991,  by Blamire et al\cite{BlamirePRL1991}. Eliash'berg made clear in 1969 that by a stationary absorption of microwaves the quasiparticles reach a nonequilibrium distribution $f(E)$ in which low-lying states are being depopulated and higher energy states become more densely populated (Fig.~\ref{Fig3}). The many quasiparticles, just above the energy-gap, where the density-of states is high, absorb the energy of the microwaves and shifts them to $\Delta+\hbar\omega$, assuming the energy-relaxation is slow enough to create the stationary non-equilibrium distribution, as shown in Fig.~\ref{Fig3}.  (Eliash'berg\cite{Eliashberg1970} assumed an energy-independent relaxation-rate, which was later replaced by the full expression for electron-phonon relaxation by Chang and Scalapino\cite{ChangScalapino1977}, which has been used in the calculations shown in Fig.~\ref{Fig3}.) At, for example a temperature of $320~mK$ the occupation is substantially reduced. Because the low-lying states weigh more heavily in the BCS gap-equation,       
\begin{equation}
\Delta=\lambda\int^{\hbar\omega_D}_\Delta\frac{\Delta}{\sqrt{E^2-\Delta^2}}\left[1-2f(E)\right]
\label{Gapequation}
\end{equation} 
the system appears 'cooler' and the energy gap is enhanced, and even continues to be present beyond the equilibrium $T_c$. In Eq.~\ref{Gapequation}, $\lambda$ is the electron-phonon coupling constant, and $f(E)$ is the distribution-function, which in equilibrium is the thermal Fermi-Dirac distribution, also shown in Fig.~\ref{Fig3}. 

\begin{figure}[t!]
   	\includegraphics[width=1\linewidth, angle=-90] {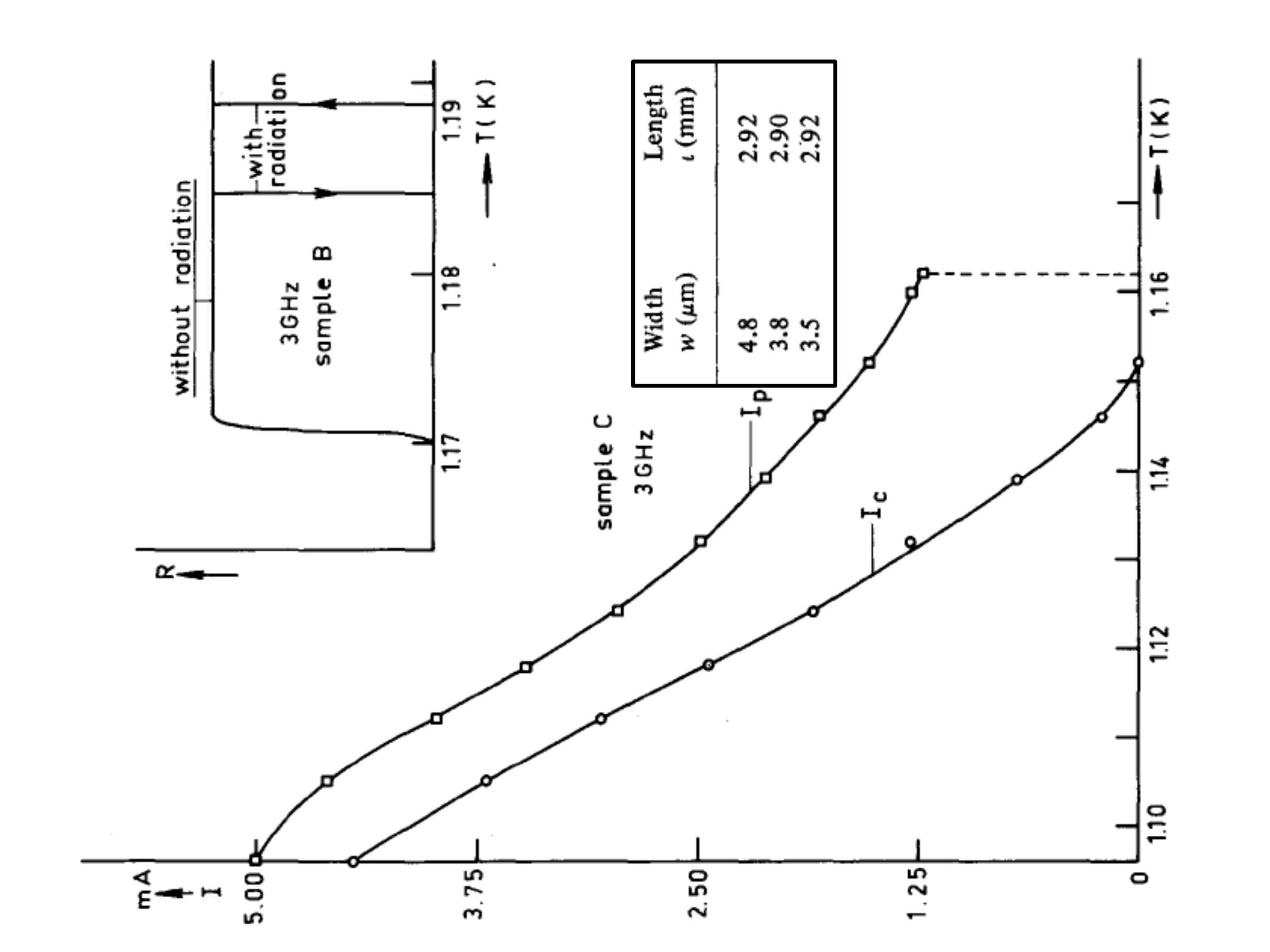}   		
	\caption{Microwave-enhancement of the critical pair-breaking current in about $4~\mu m$ wide and very long, 3 mm, aluminium strips. In addition the inset shows the resistive transition with and without microwaves (Taken from Klapwijk and Mooij\cite{KlapwijkMooij1976})}
   	\label{Fig4}
   \end{figure}

The gradual empirical disappearance of the microwave-enhanced critical current in superconducting microbridges suggested that it was not a property of the Josephson-effect, which made the theory of Eliash'berg a separate problem signaling the response of a superconductor to the microwave field. The theory was built to calculate the superconducting energy gap, a quantity which in principle could be measured with a tunnel-junction. Avoiding the challenge to develop also the tunnel-junction technology Klapwijk and Mooij\cite{KlapwijkMooij1976} studied the critical current of long, narrow strips of the superconductor aluminium. They argued that if this critical current, which ought to be the intrinsic critical pair-breaking current,  would increase it would be a clear demonstration of an increase of the energy-gap of the superconductor. Moreover, the scratching apparatus, developed for the short microbridges was ideally suited for this task as well. Convincing results appeared immediately (Fig.~\ref{Fig4}), including the clearcut observation of an enhancement of the critical temperature, measured as a change of the resistive transition from a second order to a first order phase transition. These results were submitted in August 1975 to Physica, who had just started a 'letter'-mode of operation and promised a fast reviewing process. Satisfied that it had been submitted  the results were presented at an informal seminar at the International Conference on Low Temperature Physics (LT14 at Helsinki). The results were also brought to the attention of Albert Schmid, pointing out that the change from second order to first order phase transition was a remarkable ingredient in the data, which he quickly addressed\cite{Schmid1977}. In the following year new experimental results were published by Latyshev and Nad'\cite{Latyshev1974,Latyshev1976}, which made the same argument about the critical pair-breaking current in long bridges, using the material tin (Sn) with a shorter coherence length. In addition, they also mention an enhancement of $T_c$.    
 
Further evidence emerged from an experiment carried out by Tredwell and Jacobsen\cite{TredwellPRL1975,TredwellPRB1976} showing enhancement of a critical current due to \textit{phonons}, rather than photons,  which is implicitly in agreement with Eliash'berg's mechanism. This independent work benefitted from close contact with Shapiro, who was informed about  the work going on at Copenhagen as well as at Kharkov. The work is written as if it is a measurement of the energy gap, although in practice it is very much a measurement of the critical current of a microbridge or a pointcontact. 

An important concern was that although the new experiments made clear that stimulation of superconductivity by microwaves, or phonons, was an effect intrinsic to the superconducting state, none of the experiments measured the energy gap $\Delta$ itself. A series of experiments\cite{Kommers1977,DahlbergOrbachSchuller1979,HallSoulen1980} with tunnel-junctions was carried out, but all of them suffered from a strong contribution of photon-assistance to the tunnelling current, which prevented a quantitative evaluation of the, sometimes, observed enhanced energy gap. The alternative route was a continued focus on the intrinsic nature of the critical pair-breaking current. Although the concept of the critical pair-breaking current was well known\cite{Bardeen1962} there were no known experimental cases in which it was directly measured, and the most commonly used analysis was not microscopic but based on the phenomenological Ginzburg-Landau equations. The modern expression\cite{Anthore2003} is based on the Usadel-equations\cite{Usadel1970} and given by:       
\begin{equation}
j_s(\vec{r})=\frac{\sigma}{e}\int^{\infty}_0 dE \left[1-2f\left({E,T}\right)\right]\Im (\sin^2\theta)[\vec{\nabla}\phi-\frac{2e}{\hbar}\vec{ A}]
\label{supercurrent}
\end{equation}  
in which $\theta$ is the energy-dependent pairing angle expressing the strength of the superconducting state and $f(E)$ the distribution function. It is unlike the Ginzburg-Landau equations valid for all temperatures. Experimentally,  the application of this type of theory to a measurement of the critical pair-breaking current was carried out by Romijn et al\cite{Romijn1982} with convincing results. A measurement of the predicted modifications of the density of states, or the quantity $\theta$ was carried out by Anthore et al\cite{Anthore2003}. Therefore there is little doubt that a measurement of an increase of the critical pair-breaking current by microwaves needs to be understood as a manifestation of a change in $f(E)$ as shown in Fig. \ref{Fig3}. The microwave-modified distribution function, inserted in Eq. \ref{Gapequation} solved together with Eq.~\ref{supercurrent}, which on its turn also depends on $\Delta$ and the distribution-function $f(E)$, determine together the observed critical pair breaking current. The measurement of the resistive transition, shown in the inset of Fig.~\ref{Fig4}, is in itself a striking result, which caused some additional questions\cite{FalcoSSC1980}, but appears to have ended with an exclamation mark\cite{MooijSSC1980}. A proper description of the microwave-induced \textit{emergence of superconductivity out of the normal state} remains to be developed (see Section \ref{Seven}).

A remaining question is why an increased critical current was initially observed by both Wyatt and Dayem, and why it disappeared from the experimental and theoretical field of view of research on Josephson-junctions. The additional question is whether there are experiments, which reveal better the physics that Wyatt and Dayem had accidentally stumbled upon. The answer is that the relevant physics observed by Wyatt and Dayem can only be called the Eliash'berg effect, if it is controlled by the phenomenon of energy-gap-enhancement, intrinsic to a superconducting film in a microwave field, and hence, is disconnected from the phase-coherence dominated Josephson-effect. The Eliash'berg-effect reflects the sensitivity of the superconducting state to the distribution-function $f(E)$. The essential idea of the Eliash'berg effect is that the superconducting energy gap can be increased by changing the distribution $f(E)$ from the equilibrium version (cf. Fig. \ref{Fig3}), the Fermi-Dirac distribution around the Fermi-level $E_F$ and at the equilibrium temperature $T$ of the electron-system, towards a distribution with population-inversion. In the Eliash'berg-effect this is achieved by the application of a microwave-field. The earlier idea of a population-inversion as envisioned by Parmenter\cite{Parmenter1961} in 1961 using two tunnel-junctions in series, may have gone unnoticed for a while by the overwhelming presence of the Josephson-effect appearing in 1962. There is no doubt that the Eliash'berg effect exists. In the interpretation of the data of Wyatt et al and Dayem et al, it is still an open question whether in their constriction-type short microbridges,  the enhancement of the energy-gap in the wide electrodes dominated the response, or that they obsverved a non-equilibrium response of the Josephson-effect, \textit{i.e.} due to microwave-enhanced phase-coherence, to be discussed below.     

\section{The non-equilibrium Josephson-current in SNS-junctions}\label{Four}
In discussing short constriction-type Josephson-junctions we have focused on the case in which one material is used, and only the geometry is shaped, in fact this is the case with superfluid helium as well. Another type of metallic weak link is the SNS junction in which the superconductivity is weakened by the use of the N-part, relying on the proximity-effect for the coupling. Early work by Notarys et al\cite{NotarysJAP1973} configured them in the form of a narrow constriction, which brought the normal state resistance in a usable range of values for conventional electronics. By using the superconductor tin (Sn) and gold (Au) as a normal metal they obtained a convincing Josephson response with a supercurrent emerging from zero and evolving into a Bessel-function-like behavior with increasing power\cite{NotarysPRL1973}. It clearly signaled a strengthening of the coupling through the proximity-effect by microwaves. Similar results were reported in 1979 by  Warlaumont et al\cite{WarlaumontPRL1979} using lead-copper-lead microbridges, showing a microwave-enhanced critical current down to temperatures of $0.2~ T_c $, which is not compatible with the original Eliash'berg mechanism for the superconductor itself\cite{TikhonovPRB2018}. This is also clear from the distribution-functions in the presence of microwaves shown in Fig.~\ref{Fig3} for 120 mK, using aluminium. 

These relatively early results articulated a strong case that the story of the microwave-enhancement of the critical supercurrent was not closed with the gap-enhancement in a uniform superconductor studied by Eliash'berg. In practice, these experimental results on SNS-bridges were largely ignored. The applied theoretical framework was based on a very elementary understanding of the proximity-effect, derived from the summary by Deutscher and De Gennes\cite{ParksII1969}. The proximity-effect was understood, in the spirit of the Ginzburg-Landau theory, as an exponential decay of the order parameter over a coherence length approximately given by $\sqrt{\hbar D/k_BT}$. The missing ingredient was the energy-dependence of the states participating in the proximity-effect, as well as the phase-dependence, which eventually leads to the Josephson-current. On top of that, one needs a theory which also allows for a non-equilibrium occupation of states, similarly to Eliash'berg's theory for the conventional uniform superconducting state, but now for a proximitized system. Some progress along this path to the non-equilibrium proximity-effect was made available in an article written by Volkov and co-workers\cite{VZK1993}. A number of steps in the evolution of the needed theory have been presented in a review by Belzig et al\cite{Belzig1999}. One of the authors has summarized a number of relevant experimental results, in a review, which connects the proximity-effect to Andreev-reflections\cite{Klapwijk2004}. Currently ongoing research on hybrid nanostructures are variations on the theme of the 'proximity-effect under nonequilibrium conditions',  a program already called for in principle by the experiments of Notarys et al\cite{NotarysPRL1973} and Warlaumont et al\cite{WarlaumontPRL1979} on microwave-enhanced Josephson-currents in proximity-effect microbridges. 
\begin{figure}[t!]
   	\includegraphics[width=1\linewidth, angle=0] {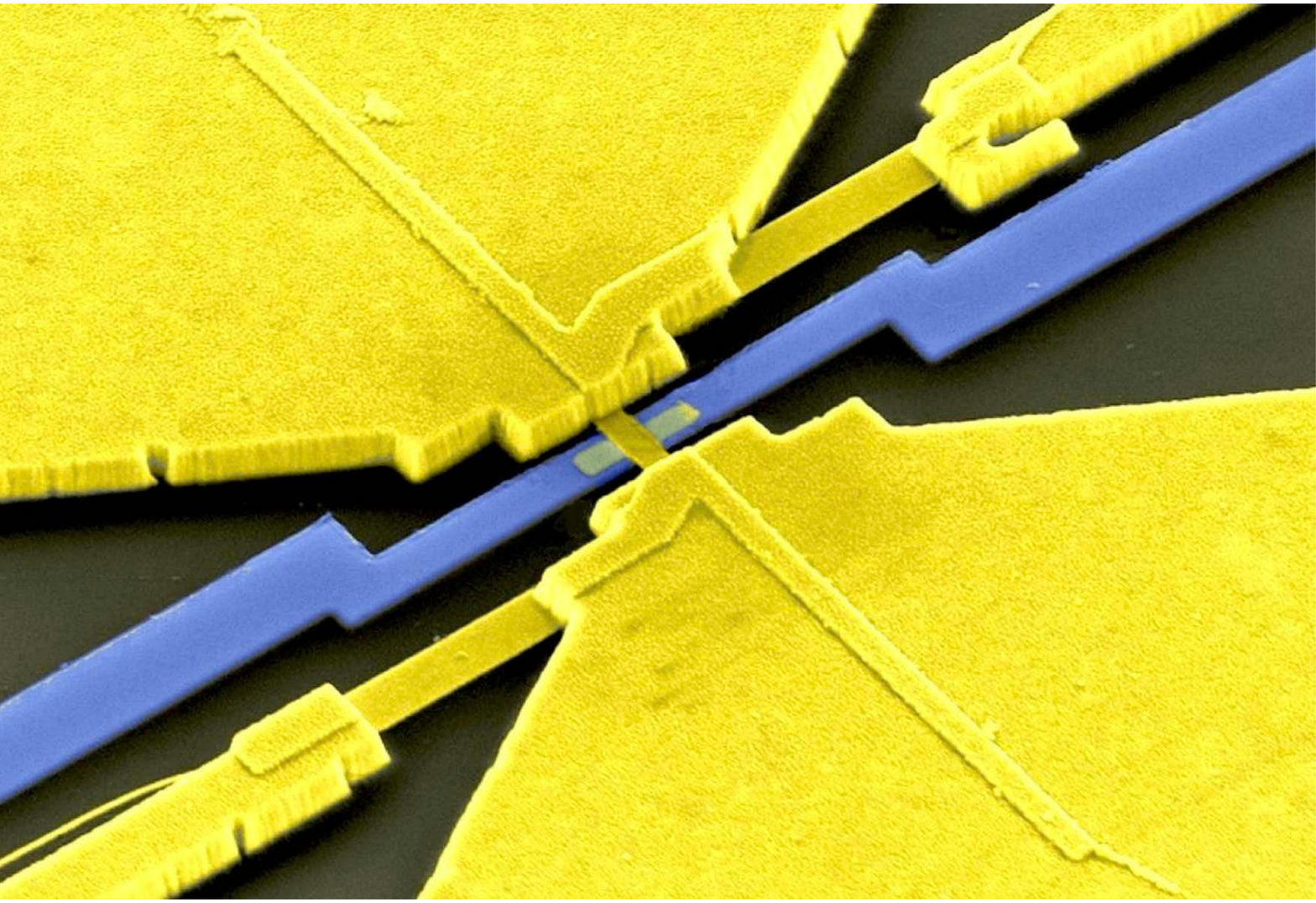}   		
	\caption{SEM-image of a SNS-type microbridge, using superconducting electrodes (niobium, colored blue), and connected by a 100 nm wide normal metal (gold, colored yellow), which is connected to large normal metal reservoirs to supply a nonequilibrium occupation of states to the proximitized normal center-part. (Taken from Baselmans et al\cite{BaselmansPRB2001})}
   	\label{Fig5}
   \end{figure}

An intriguing aspect of the current understanding of SNS type nanostructures is to probe experimentally the interplay between the phase- and energy-dependent density of states and the distribution-function, $f(E)$. For microwaves there are no new experiments available but related experiments have been carried out by Morpurgo et al\cite{Morpurgo1998} and Baselmans et al.\cite{BaselmansNature1998,BaselmansPRL2002}. The experiments are based on Eq.~\ref{supercurrent}. The pairing angle $\theta$ is for an SNS-junction representative of the proximity-effect. It is energy-dependent for $E<\Delta$ with $\Delta$ the energy gap of the superconducting electrodes, and also dependent on the phase-difference between the two superconductors $\phi=\phi_L-\phi_R$. In equilibrium $f(E)$ is the Fermi-Dirac distribution-function at the bath temperature $T$ and one finds the critical current for a SNS-junction. The experimental configuration used is shown in Fig.~\ref{Fig5}. In one experiment\cite{Morpurgo1998} the N-part of the SNS junction is connected with a relatively long normal metal wire through which a current is applied. If the wire is long enough a parabolic temperature profile for the distribution of the electrons is created with an effective temperature at the center of $T_{eff}=\sqrt{T^2+(aV)^2}$, with $T$ the bath temperature and $a$ is $3.2~K/{mV}$, as quantified by Kozub et al\cite{Kozub1995}. As expected with increasing temperature in N the critical current of the SNS-junction goes down. The more interesting case is when the distribution of electrons in the normal wire is non-thermal, studied in detail by Pothier et al\cite{Pothier1997} for shorter wires in which the electron-electron interaction rate is not capable of creating a thermal distribution. One finds that $f(E)$ becomes a two step distribution function, which marks  the difference between the two distributions of the contacts, each at a different voltage. By applying such a non-equilibrium distribution to the energy-dependent density of states of the proximitized N-part of the SNS junction, Baselmans et al\cite{BaselmansNature1998,BaselmansPRL2002} showed that the conventional $\sin\phi$ dependence of a Josephson-junction reverses sign and becomes a $\pi$-junction.  The deliberate choice of a non-thermal $f(E)$ together with the energy-dependence of the proximity-effect causes this result. It is a very clear example of the non-equilibrium proximity-effect.             

Another example, which is currently much less clear, is the voltage-carrying state of an SNS-junction. If one exceeds the critical current, the RSJ-model provides the elementary framework of an oscillating voltage at the Josephson-frequency, which leads to a time-averaged DC current-voltage characteristic. The details depend on the impedance of the environment. Microscopically, if we assume a voltage-bias the proximity-induced density of states, which depends on the phase difference $\phi$, will become an oscillatory quantity at the Josephson-frequency, which is determining the oscillatory supercurrent. The actual value of this oscillating supercurrent will depend on the occupation  of the rapidly changing density of states. In early research on microbridges it was found that the critical current, the amplitude in the $I_c\sin\phi$-relation \textit{increases} from the value at $V=0$ to reach a maximum when $\omega\tau_{inel}\approx 1$. The reason is that the Josephson-frequency is so high that there is no time to reach thermal equilibrium for the distribution of electrons over the energies. This dynamic enhancement effect has been discussed by Schmid et al\cite{SST1980}, but it has to the best of our knowledge not been discussed in the context of SNS-junctions and modern versions of it. It is an important topic because it is obvious that at finite voltage power is fed into the system\cite{Divin1974}, which potentially reduces the amplitude of the oscillating Josephson-current. This effect led to the development of variable-thickness bridges as discussed in Section \ref{Two}. In recent years the topic of hot electrons has been refreshed in the application to SNS-systems in the strongly driven regime by Courtois et al\cite{Courtois2008}, followed by a conjecture that this local temperature affects the observation of Shapiro-steps\cite{DeCeccoCourtois2016}, a discussion reminiscent of research in the 70-ies of Octavio et al\cite{Octavio1977} and Tinkham et al\cite{Tinkham1977}. However, an important difference with the current generation of devices, operating in the quantum transport regime of Landauer, is that they can not be described by a local temperature, but are by definition working with a non-equilibrium occupation of states in the scattering regime (Section \ref{Five}).     

In recent theoretical work by Virtanen et al\cite{Virtanen2010} the application of microwave radiation to SNS-junctions has been readdressed, and compared with experiments carried out by Chiodi et al\cite{ChiodiPRL2009} and Fuechsle et al\cite{FuechslePRL2009}. It brings together measurements on the enhanced critical current as well as on the current-phase relationship as modified by the occupation of states by the microwaves. It also brings an answer to the old question posed by the work of Notarys et al\cite{NotarysPRL1973} and Warlaumont et al\cite{WarlaumontPRL1979}, why a microwave-enhanced critical current is observed in a system in which the original Eliashberg-mechanism for the superconductor itself can not be at work. Virtanen et al apply the modern energy- and phase-dependent description of the proximity-effect in SNS together with a redistribution of the electrons over the energies in the spirit of Eliash'berg, but applied to a proximitized SNS system. In both cases the $V=0$ properties are measured and calculated. The dynamic aspects have also been addressed in recent work by Chiodi et al\cite{ChiodiSR2011} with related theoretical work by Tikhonov and Feigel'man\cite{TikhonovPRB2015}. The recent work on the nonequilibrium proximity-effect in SNS-junctions makes clear that a nonequilibrium distribution $f(E)$ for the N-part deeply affects the static and the dynamic properties of the SNS-Josephson junction, which is quite different from raising the temperature. This is also the essential message of Eliash'berg's original idea about microwave-stimulated superconductivity. It can manifest itself as a possible enhancement of the energy gap, but also as an enhancement of the Josephson-coupling.                

\section{Superconducting Landauer-type quantum pointcontacts}\label{Five}
   
The pioneering experiments on nanoscale superconducting devices, sketched in Section \ref{Two}, evolved in the 80-ies quite naturally to mesoscopic physics in normal metals and semiconductors. The emphasis was on transport in artificially made structures with a length shorter than the single particle phase coherence length\cite{BeenakkerVanHouten1991}, in many cases identical to the inelastic scattering length. It defined the regime of quantum coherent transport in the normal state.  Following insight from Landauer it leads for example to an expression for the two-point conductance\cite{NazarovBlanter2009} of: 
\begin{equation}
I=\frac{G_Q}{e}\sum_n \int dE~T_n(E)\left[f_L(E)-f_R(E)\right]
\label{Landauer}
\end{equation} 
with $G_Q=e^2/{\pi\hbar}$, the sum over $n$ the number of modes available for transport, and the integral over the energies. In this case, unlike in Eq.~\ref{supercurrent}, the distribution functions $f_L$ and $f_R$ are spatially separated, indicated by the subscripts $L$ for left and $R$ for right. It is furthermore assumed that the nanostructure is characterized by a set of transmission coefficients $T_n$ and is connected to equilibrium reservoirs, which emit or absorb electron waves. The conductance is measured by giving the two reservoirs a different potential $eV$, which means that electrons with a higher energy emitted from $L$ absorbed at $R$. This might in principle lead to an increase in temperature of the right reservoir. However, the reservoirs are assumed to be large and therefore this effect is considered to be negligible, unless, in practice, the geometry is unfavorable.  

\begin{figure*}[t!]
   	\includegraphics[width=1\linewidth] {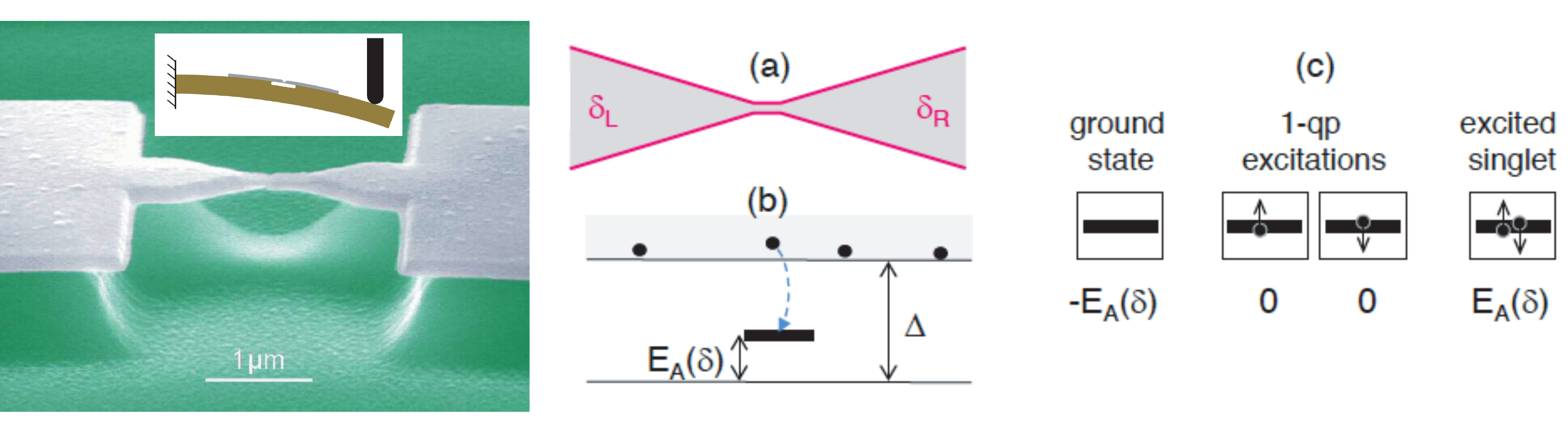}   		
	\caption{SEM-picture of a mechanical break junction, which leads to atomic scale contacts between two superconductors of the same material, with the important advantage of being a tunable contact. The supercurrent is carried by Andreev bound states with energy $-E_A(\delta)$, which depends on the phase-difference $\delta=\delta_L-\delta_R$. The twofold discrete energy-levels can be selectively populated determining the strength of the Josephson-coupling. (Reproduced from Bretheau et al\cite{Bretheau2012})}
   	\label{Fig6}
   \end{figure*}

Although Eq.~\ref{Landauer} is for normal metals it is worth to compare this approach with the point of view used for non-equilibrium superconductivity. In the latter case, we assume a local superconducting density of states, which may be the result of an inhomogeneous system such as a SNS junction, and a local distribution-function $f(E)$. This local distribution function could have an enhanced electron temperature or be non-thermal. In both cases it will affect the supercurrent. In understanding the properties of these diffusive SNS-junctions,  the challenge is to determine under driven conditions the local $f(E)$. In the Landauer point of view two distribution-functions play a role, both from an equilibrium reservoir, separate from the scattering region. The occupation of the states in the scattering region are fully determined by the difference between $f_L(E)$ and $f_R(E)$ each at a different voltage. The Landauer-system is assumed to separate spatially the reservoirs and the scattering region, and there is no need to assign a non-equilibrium distribution-function or a temperature to the scattering region. It is for any driven case locally a non-equilibrium region, and since there is no inelastic scattering assumed, it is always non-thermal.  

\begin{figure*}[t!]
   	\includegraphics[width=1\linewidth] {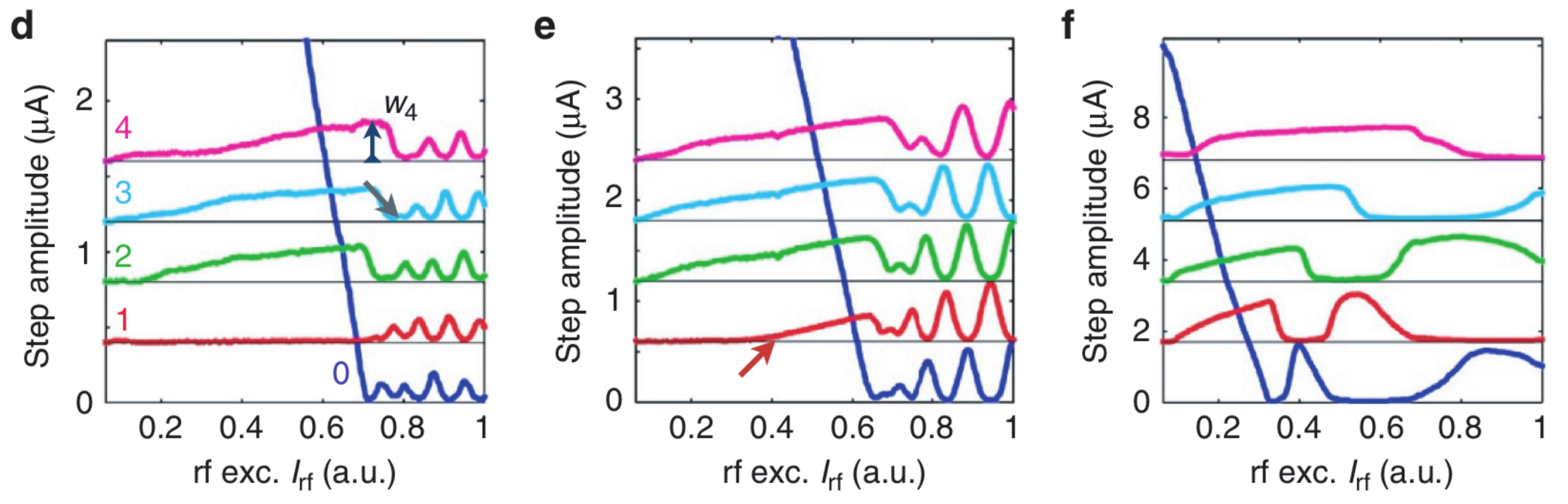}   		
	\caption{Shapiro -step measurements for Josephson-junctions based on the topological materials strained HgTe. The heights of the Shapiro-steps as a function of microwave-amplitude  for 2.7 GHz (d), 5.3 GHz (e) and 11.2 GHz (f) are anomalous. Note the absence of steps at various values of $I_{rf}$. (Taken from Wiedenmann et al\cite{Wiedenmann2016})}
   	\label{Fig7}
   \end{figure*}

Beenakker\cite{Beenakker1991} has shown that in the short-channel limit the current is carried by Andreev bound states (Fig. \ref{Fig7}) of the form
\begin{equation}
E(\phi)=\sum_{p=1}^N E_p(\phi)=\Delta \sum_{p=1}^N \sqrt{1-T_p\sin^2(\phi/2)} 
\label{ABS}
\end{equation}   
with $T_p$ the transmission coefficient for channel $p$, assuming that more than one channel contributes. The Andreev bound states take the role of pairing angle $\theta$ used in Eq.~\ref{supercurrent}. Under these conditions the supercurrent in such a Landauer pointcontact is given by
\begin{equation}
I(\phi)=\frac{e\Delta^2}{2\hbar} \sum_{p=1}^N \frac{T_p\sin\phi}{1-T_p\sin^2(\phi/2)}\left[1-2f(E_p)\right]
\label{QuantumSupercurrent}
\end{equation}       
assuming a scattering region connected to superconductors with energy gap $\Delta$, and $f(E)$ the equilibrium Fermi-Dirac distribution, the same meaning as in Eq.~\ref{Landauer}, but for zero applied voltage. This expression is for $T_p=1$ in agreement with the calculation of the critical current by Kulik and Omel'yanchuk\cite{KOII1977} based on the Eilenberger equations. In considering non-equilibrium effects it is clear that for example the temperature dependence is contained in the Fermi-Dirac distribution and in the value of the energy gap $\Delta$, both of which are boundary conditions of the scattering problem. If a voltage is applied the phase-difference evolves through the Josephson-relation $\partial\phi/{\partial t}=2eV/\hbar$, while at the same time the continuum of states with $E>\Delta$ will be populated in a way analogous to Eq.~\ref{Landauer}. The population of the Andreev bound state, and hence the supercurrent, will depend on the exchange between the non-thermal occupation of the states in the continuum.  Obviously, this is not an easy problem, although it is reminiscent of the dynamic enhancement discussed in Section \ref{Four}. Recently, a generalized Landauer-formula for the superconducting quantum pointcontacts was proposed by Glazman and co-workers\cite{Glazman2019} to include dissipation.  

The original microbridge\cite{AndersonDayem1964}, a constriction in a superconducting thin-film, evolved further into what is called a mechanical break-junction (Fig.~\ref{Fig6}). It uses the progress in lithography and the gentle mechanical placement emerging from scanning tunnelling microscopes. First, the wire is broken mechanically at a lithographically created constriction in a superconducting metal, followed by a subsequent femtometer-accuracy mechanical approach. This technique permits the study of single-atom superconducting contacts, all made of the same material, with an adjustable weak link consisting of contacts between atomic orbitals of two atoms, where the irregularities of the two broken surfaces happen to touch first. In this case, the two superconducting films are bulk-electrodes, which touch on the scale of a single atom. They allow the determination of the contribution of the different modes to the supercurrent-transport and for each of them the transmissivity\cite{Chauvin2007}. Knowing this information one can easily infer what the supercurrent as a function of the phase-difference  will be, which has been found to be in excellent agreement with the experimental results by Della Rocca et al\cite{DellaRocca2007}, showing that $\sin\phi$ is not a universal characteristic of the Josephson-effect, but dependent on the number of modes and their transmissivity. In addition, the same group has combined this with a controlled application of microwaves in order to carry out supercurrent spectroscopy of the Andreev-levels\cite{Bretheau2013}.

The Landauer-type superconducting point-contacts at finite voltage play an interesting and challenging role in recent work on superconducting weak links of topological materials.  For weak links of these materials it is to be expected that apart from the conventional Andreev bound states, such as given in Eq.~\ref{ABS} and Fig.~\ref{Fig6} gapless Andreev bound states occur giving rise to a supercurrent of $4\pi$-periodicity.  One way to search for this periodicity is by studying Shapiro-steps, because for  $4\pi$ periodicity one expects only steps at even integers. Fig.~\ref{Fig7} shows a selection of step heights of Shapiro-steps observed in the current-voltage characteristics of niobium-strained HgTe-niobium devices, for 3 different frequencies (2.7, 5.3 and 11.2 GHz). In these data\cite{Wiedenmann2016} the focus was on the absence of a step at $n=1$, which was taken as an indication of the absence of odd-steps. In subsequent work\cite{Bocquillon2016,Deacon2017} using a HgTe quantum well a much clearer absence of the odd series was found at the lower end of the used frequencies (6.6, 3.5, 1.8, 1.0 and 0.8 GHz). This indicates that the $I(\phi)$-relation is not periodic in $2\pi$, as implied in Eq.~\ref{QuantumSupercurrent} but in $4\pi$. This is an indication of Andreev bound-states different from Eq.~\ref{ABS} but rather of gapless states. An interesting question is whether the frequency-dependence of the observability of this evidence for $4\pi$-periodicity is related to the dynamics of the relevant processes.  Additionally a striking effect in the data of Fig.~\ref{Fig7} is not only the absence of odd steps, but also the large ranges of the $I_{rf}$-amplitude over which they appear to be absent before to reappear.  These are all clear deviations from the conventional analysis of the RSJ-model. Dominguez et al\cite{Dominguez2017,PicoCortesPRB2017} have tried to reconcile the observations by assuming the coexistence of a $2\pi$ and a $4\pi$ periodic contribution to the supercurrent. 
This phenomenological approach has many unknown parameters and leaves out of sight the microscopic processes. It is clear that in the presence of a voltage $V$, the phase difference evolves in time with the Josephson relation, $\partial\phi/{\partial t}=2eV/\hbar$. It means that the energies $E$ of the Andreev bound states, Eq.~\ref{ABS},  evolve in time and also the supercurrent (Eq.~\ref{QuantumSupercurrent}). The same is true for the conjectured gapless Andreev bound states. In this dynamic process we need to keep track of the occupation of these states as well. In short, understanding the details of the data shown in Fig.~\ref{Fig7} continue to pose an interesting challenge.    

Finally,  in closing this Section in connection to the microwave-stimulated superconductivity we like to draw the attention to recent work of Bergeret and co-workers\cite{Bergeret2010}. They have shown theoretically that in a superconducting quantum point contact at finite temperature when the lowest Andreev bound state is not fully occupied a microwave-enhanced supercurrent can be obtained by applying microwave-frequencies which take into account the energy of the Andreev bound states. It is again a means of selectively populating and depopulating the supercurrent-carrying  states, which is the basis of the supercurrent spectroscopy by microwaves, carried out by Bretheau et al\cite{Bretheau2013}.

\section{Eliash'berg effect in superconducting resonators}\label{Six}

The most important contribution of the Eliash'berg-effect to superconductivity and the microwave-enhancement is the emphasis on the role of the distribution-function $f(E)$ on the temperature-dependence of the energy gap $\Delta$. By removing quasiparticles from the edge of the Fermi-sphere the critical temperature $T_c$ increases, which is readily understood based on the microscopic theory of superconductivity. 

Even so, it continues to be counterintuitive, as illustrated by recent work on superconducting microwave resonators\cite{DeVisserPRL2014}. These experiments were carried out at temperatures between $T_c/10$ and $T_c/3$, much lower than experiments on critical current and gap enhancement. At intermediate temperatures, with increasing microwave power an increase in the resonant frequency shows up, indicative of a lower kinetic inductance, which goes together with a strong reduction in microwave losses, which can both be explained by a strong non-equilibrium $f(E)$ (Fig.~\ref{Fig3} bottom panel). The effect disappears, as expected, at the lowest temperatures when there are no longer thermal quasiparticles available. Interestingly, at these low temperatures there is still a response to the microwave power observed, which could be taken for some form of heating, despite of the fact that $\hbar\omega \ll 2\Delta$ and the density of quasiparticles is very low. The presence of a microwave-power-dependent number of excess quasiparticles can be understood by taking the Eliash'berg formalism and to solve the steady state $f(E)$ under microwave absorption, the results of which are shown in Fig.~\ref{Fig3} (top panel). A strong non-equilibrium $f(E)$ can build up due to the slow scattering and recombination processes at low temperature. However, the question remains how the very few quasiparticles at $T_c/10$ can start to build up this non-equilibrium distribution. The explanation is that Eliash'berg had focused in his analysis on the absorption by quasiparticles and ignored the response of the superconducting condensate to the microwave field. Similar to the effect of a dc current on the properties of the superconducting ground state\cite{Anthore2003} one can include the effects of an ac current \cite{Semenov2016, Semenov2019}. 

The lesson to be drawn from Eliash'berg's theory is that there is no such thing as simple heating in superconductivity, except by really raising the temperature. In all other cases it is changing the distribution of electrons over the energies $f(E)$, with its rich panorama of consequences.  

\section{Re-emergence of photo-stimulated superconductivity}\label{Seven}
The present reconstruction of the fate of the Dayem-Wyatt and the Eliash'berg effect raises a couple of questions. One of them is the fact that it was an active field of research, starting around 1970 and then died around 1984. Why? The main reason was that the effect showed up in the search for the Josephson-effect in superconducting microbridges. In the early 80-ies the interest in microbridges and pointcontacts gradually disappeared, because they were not useful for detection of radiation and the interest shifted to tunnel-junctions, as had happened for SQUIDS a bit earlier. Another reason was that research-interest shifted towards the emerging field of mesoscopic physics in normal metals and semiconductors, supported by the increasing presence of clean room technology. A third reason was the disappearance of the IBM and Bell Labs superconducting computer program and the appearance of high Tc superconductivity, which both were definitely disruptive events.   

The irony is that even today a thorough quantitative experimental evaluation of the Eliash'berg effect has not been carried out. In one experiment a superconducting cylinder has been studied to measure the order parameter\cite{PalsDobben1980}. In principle, one would like to measure in a reliable way the superconducting energy gap in the presence of a microwave field in a uniform system and without affecting the measuring process. A few experiments have been carried out with tunnel-junctions\cite{Kommers1977,DahlbergOrbachSchuller1979,HallSoulen1980}, but they are strongly influenced by the presence of photon-assisted tunneling. Only one experiment has circumvented this problem\cite{HorstmanWolter1981,WolterHorstman1981}, but the results are primarily qualitative. In principle the present level of microwave-technology has the potential to carry out a much better quantitative study\cite{Semenov2019}. In all other cases a microbridge-type device was studied in which, as discussed, the response is not due to gap-enhancement, but due an enhancement of the phase-coherence, described in Sections \ref{Four} and \ref{Five}.  
                         
A very important, but insufficiently studied problem of the Eliash'berg-effect is the emergence of superconductivity out of the normal state. In principle, Eliash'berg's theory starts with  the existence of an energy-gap. Experimentally, one finds that if the normal metal is exposed to a microwave field, the normal metal jumps to the superconducting state\cite{KlapwijkMooij1976}, which qualitatively can be understood based on the analysis of Schmid\cite{Schmid1977}. For a material like aluminium or tin above the equilibrium critical temperature, $T_c$, we might assume superconducting fluctuations, usually distinguished in Aslamazov-Larkin and Maki-Thomson contributions\cite{MooijSSC1980}. These fluctuations could act as the nucleation-centers from which superconductivity emerges due to the application of a microwave-field. Whether this emergence is driven by enhanced phase-coherence, or by an increase of the amplitude of the superconducting fluctuations is not known. Similar questions are currently being raised by the recent work on light-induced superconductivity\cite{Cavalleri2017} and transient enhancement of superconductivity\cite{Fausti2011,MitranoNature2016,Demsar2016}, which also stimulated work on strongly disordered superconductors such as NbN\cite{BeckPRL2013}. The present review makes the distinction between gap-enhancement in uniform superconductors and enhanced phase-coherence in Josephson-coupled weak links, which is likely to be relevant here as well. Additionally, a lattice-induced change of the pairing-interaction is a third possibility. One crucial question in studying the response to radiation is whether one observes an increased Josephson coupling strength, without a change of the pairing gap, or a change in the energy gap, without a change in the Josephson-coupling strength. In many recent cases, the experimental data are analysed by disregarding the Eliashberg mechanism for a uniform superconductor, without reflecting upon the effect of radiation on the Josephson-coupling. This subject is also highly relevant for optical studies of strongly disordered materials\cite{BeckPRL2013}.  It suggests potentially interesting research on the electrodynamics of mesoscopic arrays as model-systems for superconductivity in complex materials\cite{Han14,SacepeReview2020}. The early experimental results of Notarys et al\cite{NotarysPRL1973} and Warlaumont et al\cite{WarlaumontPRL1979} are a clear demonstration of radiation-stimulated Josephson-coupling, which from the point of view of complex materials may be understood as the emergence of superconductivity out of a metallic state.      
\section*{Acknowledgments} We like to acknowledge useful input from P.E. Lindelof, F.S. Bergeret, H. Pothier, A.D. Caviglia, and two anonymous reviewers, who helped us to improve the manuscript. TMK acknowledges financial support from the Russian Scientific Foundation, Grant No. 17-72-30036 and from the W\"urzburg-Dresden Center of Excellence on Complexity and topology in quantum matter (CT.QMAT). PJdV is financially supported by the Netherlands Organisation for Scientific Research NWO (Veni Grant No. 639.041.750 and Projectruimte 680-91-127)


\begin{thebibliography}{104}%
\makeatletter
\providecommand \@ifxundefined [1]{%
 \@ifx{#1\undefined}
}%
\providecommand \@ifnum [1]{%
 \ifnum #1\expandafter \@firstoftwo
 \else \expandafter \@secondoftwo
 \fi
}%
\providecommand \@ifx [1]{%
 \ifx #1\expandafter \@firstoftwo
 \else \expandafter \@secondoftwo
 \fi
}%
\providecommand \natexlab [1]{#1}%
\providecommand \enquote  [1]{``#1''}%
\providecommand \bibnamefont  [1]{#1}%
\providecommand \bibfnamefont [1]{#1}%
\providecommand \citenamefont [1]{#1}%
\providecommand \href@noop [0]{\@secondoftwo}%
\providecommand \href [0]{\begingroup \@sanitize@url \@href}%
\providecommand \@href[1]{\@@startlink{#1}\@@href}%
\providecommand \@@href[1]{\endgroup#1\@@endlink}%
\providecommand \@sanitize@url [0]{\catcode `\\12\catcode `\$12\catcode
  `\&12\catcode `\#12\catcode `\^12\catcode `\_12\catcode `\%12\relax}%
\providecommand \@@startlink[1]{}%
\providecommand \@@endlink[0]{}%
\providecommand \url  [0]{\begingroup\@sanitize@url \@url }%
\providecommand \@url [1]{\endgroup\@href {#1}{\urlprefix }}%
\providecommand \urlprefix  [0]{URL }%
\providecommand \Eprint [0]{\href }%
\providecommand \doibase [0]{http://dx.doi.org/}%
\providecommand \selectlanguage [0]{\@gobble}%
\providecommand \bibinfo  [0]{\@secondoftwo}%
\providecommand \bibfield  [0]{\@secondoftwo}%
\providecommand \translation [1]{[#1]}%
\providecommand \BibitemOpen [0]{}%
\providecommand \bibitemStop [0]{}%
\providecommand \bibitemNoStop [0]{.\EOS\space}%
\providecommand \EOS [0]{\spacefactor3000\relax}%
\providecommand \BibitemShut  [1]{\csname bibitem#1\endcsname}%
\let\auto@bib@innerbib\@empty
\bibitem [{\citenamefont {Anderson}\ and\ \citenamefont
  {Dayem}(1964)}]{AndersonDayem1964}%
  \BibitemOpen
  \bibfield  {author} {\bibinfo {author} {\bibfnamefont {P.~W.}\ \bibnamefont
  {Anderson}}\ and\ \bibinfo {author} {\bibfnamefont {A.~H.}\ \bibnamefont
  {Dayem}},\ }\href@noop {} {\bibfield  {journal} {\bibinfo  {journal} {Phys.
  Rev. Lett.}\ }\textbf {\bibinfo {volume} {13}},\ \bibinfo {pages} {195}
  (\bibinfo {year} {1964})}\BibitemShut {NoStop}%
\bibitem [{\citenamefont {Shapiro}(1963)}]{Shapiro1963}%
  \BibitemOpen
  \bibfield  {author} {\bibinfo {author} {\bibfnamefont {S.}~\bibnamefont
  {Shapiro}},\ }\href@noop {} {\bibfield  {journal} {\bibinfo  {journal} {Phys.
  Rev. Lett.}\ }\textbf {\bibinfo {volume} {11}},\ \bibinfo {pages} {80}
  (\bibinfo {year} {1963})}\BibitemShut {NoStop}%
\bibitem [{\citenamefont {Richards}\ and\ \citenamefont
  {Anderson}(1965)}]{Richards1965}%
  \BibitemOpen
  \bibfield  {author} {\bibinfo {author} {\bibfnamefont {P.~L.}\ \bibnamefont
  {Richards}}\ and\ \bibinfo {author} {\bibfnamefont {P.~W.}\ \bibnamefont
  {Anderson}},\ }\href@noop {} {\bibfield  {journal} {\bibinfo  {journal}
  {Phys. Rev. Lett.}\ }\textbf {\bibinfo {volume} {14}},\ \bibinfo {pages}
  {540} (\bibinfo {year} {1965})}\BibitemShut {NoStop}%
\bibitem [{\citenamefont {Varoquaux}(2015)}]{Varoquaux2015}%
  \BibitemOpen
  \bibfield  {author} {\bibinfo {author} {\bibfnamefont {E.}~\bibnamefont
  {Varoquaux}},\ }\href@noop {} {\bibfield  {journal} {\bibinfo  {journal}
  {Rev. Mod. Phys.}\ }\textbf {\bibinfo {volume} {87}},\ \bibinfo {pages} {803}
  (\bibinfo {year} {2015})}\BibitemShut {NoStop}%
\bibitem [{\citenamefont {Dayem}\ and\ \citenamefont
  {Wiegand}(1967)}]{Dayem1967}%
  \BibitemOpen
  \bibfield  {author} {\bibinfo {author} {\bibfnamefont {A.~H.}\ \bibnamefont
  {Dayem}}\ and\ \bibinfo {author} {\bibfnamefont {J.~J.}\ \bibnamefont
  {Wiegand}},\ }\href@noop {} {\bibfield  {journal} {\bibinfo  {journal} {Phys.
  Rev.}\ }\textbf {\bibinfo {volume} {155}},\ \bibinfo {pages} {419} (\bibinfo
  {year} {1967})}\BibitemShut {NoStop}%
\bibitem [{\citenamefont {Shapiro}\ \emph {et~al.}(1964)\citenamefont
  {Shapiro}, \citenamefont {Janus},\ and\ \citenamefont {Holly}}]{Shapiro1964}%
  \BibitemOpen
  \bibfield  {author} {\bibinfo {author} {\bibfnamefont {S.}~\bibnamefont
  {Shapiro}}, \bibinfo {author} {\bibfnamefont {A.~R.}\ \bibnamefont {Janus}},
  \ and\ \bibinfo {author} {\bibfnamefont {S.}~\bibnamefont {Holly}},\
  }\href@noop {} {\bibfield  {journal} {\bibinfo  {journal} {Rev. Mod. Phys.}\
  }\textbf {\bibinfo {volume} {36}},\ \bibinfo {pages} {223} (\bibinfo {year}
  {1964})}\BibitemShut {NoStop}%
\bibitem [{\citenamefont {Wyatt}\ \emph {et~al.}(1966)\citenamefont {Wyatt},
  \citenamefont {Dmitriev}, \citenamefont {Moore},\ and\ \citenamefont
  {Sheard}}]{Wyatt1966}%
  \BibitemOpen
  \bibfield  {author} {\bibinfo {author} {\bibfnamefont {A.~F.~G.}\
  \bibnamefont {Wyatt}}, \bibinfo {author} {\bibfnamefont {V.~M.}\ \bibnamefont
  {Dmitriev}}, \bibinfo {author} {\bibfnamefont {W.~S.}\ \bibnamefont {Moore}},
  \ and\ \bibinfo {author} {\bibfnamefont {F.~W.}\ \bibnamefont {Sheard}},\
  }\href@noop {} {\bibfield  {journal} {\bibinfo  {journal} {Phys. Rev. Lett.}\
  }\textbf {\bibinfo {volume} {16}},\ \bibinfo {pages} {1166} (\bibinfo {year}
  {1966})}\BibitemShut {NoStop}%
\bibitem [{\citenamefont {Grimes}\ and\ \citenamefont
  {Shapiro}(1968)}]{GrimesPRB1968}%
  \BibitemOpen
  \bibfield  {author} {\bibinfo {author} {\bibfnamefont {C.~C.}\ \bibnamefont
  {Grimes}}\ and\ \bibinfo {author} {\bibfnamefont {S.}~\bibnamefont
  {Shapiro}},\ }\href@noop {} {\bibfield  {journal} {\bibinfo  {journal} {Phys.
  Rev.}\ }\textbf {\bibinfo {volume} {169}},\ \bibinfo {pages} {397} (\bibinfo
  {year} {1968})}\BibitemShut {NoStop}%
\bibitem [{\citenamefont {Dmitriev}\ \emph {et~al.}(1971)\citenamefont
  {Dmitriev}, \citenamefont {Khristenko}, \citenamefont {Trubitsyn},\ and\
  \citenamefont {Mende}}]{Dmitriev1971}%
  \BibitemOpen
  \bibfield  {author} {\bibinfo {author} {\bibfnamefont {V.~M.}\ \bibnamefont
  {Dmitriev}}, \bibinfo {author} {\bibfnamefont {E.~V.}\ \bibnamefont
  {Khristenko}}, \bibinfo {author} {\bibfnamefont {A.~V.}\ \bibnamefont
  {Trubitsyn}}, \ and\ \bibinfo {author} {\bibfnamefont {F.~F.}\ \bibnamefont
  {Mende}},\ }\href@noop {} {\bibfield  {journal} {\bibinfo  {journal} {Ukr.
  Fiz. Zh.}\ }\textbf {\bibinfo {volume} {16}} (\bibinfo {year}
  {1971})}\BibitemShut {NoStop}%
\bibitem [{\citenamefont {G.M.Eliashberg}(1970)}]{Eliashberg1970}%
  \BibitemOpen
  \bibfield  {author} {\bibinfo {author} {\bibnamefont {G.M.Eliashberg}},\
  }\href@noop {} {\bibfield  {journal} {\bibinfo  {journal} {JETP Letters}\
  }\textbf {\bibinfo {volume} {11}},\ \bibinfo {pages} {114} (\bibinfo {year}
  {1970})}\BibitemShut {NoStop}%
\bibitem [{\citenamefont {Ivlev}(1971)}]{Ivlev1971}%
  \BibitemOpen
  \bibfield  {author} {\bibinfo {author} {\bibfnamefont {B.~I.}\ \bibnamefont
  {Ivlev}},\ }\href@noop {} {\bibfield  {journal} {\bibinfo  {journal} {Soviet
  Physics JETP}\ }\textbf {\bibinfo {volume} {32}},\ \bibinfo {pages} {563}
  (\bibinfo {year} {1971})}\BibitemShut {NoStop}%
\bibitem [{\citenamefont {Ivlev}\ and\ \citenamefont
  {Eliashberg}(1971)}]{Ivlev1971b}%
  \BibitemOpen
  \bibfield  {author} {\bibinfo {author} {\bibfnamefont {B.~I.}\ \bibnamefont
  {Ivlev}}\ and\ \bibinfo {author} {\bibfnamefont {G.~M.}\ \bibnamefont
  {Eliashberg}},\ }\href@noop {} {\bibfield  {journal} {\bibinfo  {journal}
  {JETP Letters}\ }\textbf {\bibinfo {volume} {13}},\ \bibinfo {pages} {464}
  (\bibinfo {year} {1971})}\BibitemShut {NoStop}%
\bibitem [{\citenamefont {Ivlev}\ \emph {et~al.}(1973)\citenamefont {Ivlev},
  \citenamefont {Lisitsyn},\ and\ \citenamefont {Eliashberg}}]{Ivlev1973}%
  \BibitemOpen
  \bibfield  {author} {\bibinfo {author} {\bibfnamefont {B.~I.}\ \bibnamefont
  {Ivlev}}, \bibinfo {author} {\bibfnamefont {S.~G.}\ \bibnamefont {Lisitsyn}},
  \ and\ \bibinfo {author} {\bibfnamefont {G.~M.}\ \bibnamefont {Eliashberg}},\
  }\href@noop {} {\bibfield  {journal} {\bibinfo  {journal} {J. of Low Temp
  Phys}\ }\textbf {\bibinfo {volume} {10}},\ \bibinfo {pages} {449} (\bibinfo
  {year} {1973})}\BibitemShut {NoStop}%
\bibitem [{\citenamefont {Aslamazov}\ and\ \citenamefont
  {Larkin}(1969)}]{AL1968}%
  \BibitemOpen
  \bibfield  {author} {\bibinfo {author} {\bibfnamefont {L.~G.}\ \bibnamefont
  {Aslamazov}}\ and\ \bibinfo {author} {\bibfnamefont {A.~I.}\ \bibnamefont
  {Larkin}},\ }\href@noop {} {\bibfield  {journal} {\bibinfo  {journal} {JETP
  Letters}\ }\textbf {\bibinfo {volume} {9}},\ \bibinfo {pages} {87} (\bibinfo
  {year} {(1968)1969})}\BibitemShut {NoStop}%
\bibitem [{\citenamefont {Russer}(1970)}]{Russer1970}%
  \BibitemOpen
  \bibfield  {author} {\bibinfo {author} {\bibfnamefont {P.}~\bibnamefont
  {Russer}},\ }\href@noop {} {\bibfield  {journal} {\bibinfo  {journal} {Acta
  Physica Austriaca}\ }\textbf {\bibinfo {volume} {32}},\ \bibinfo {pages}
  {373} (\bibinfo {year} {1970})}\BibitemShut {NoStop}%
\bibitem [{\citenamefont {Russer}(1972)}]{RusserJAP1972}%
  \BibitemOpen
  \bibfield  {author} {\bibinfo {author} {\bibfnamefont {P.}~\bibnamefont
  {Russer}},\ }\href@noop {} {\bibfield  {journal} {\bibinfo  {journal}
  {Journal of Applied Physics}\ }\textbf {\bibinfo {volume} {43}},\ \bibinfo
  {pages} {2008} (\bibinfo {year} {1972})}\BibitemShut {NoStop}%
\bibitem [{\citenamefont {Stewart}(1968)}]{Stewart1968}%
  \BibitemOpen
  \bibfield  {author} {\bibinfo {author} {\bibfnamefont {W.~C.}\ \bibnamefont
  {Stewart}},\ }\href@noop {} {\bibfield  {journal} {\bibinfo  {journal} {Appl.
  Phys. Lett.}\ }\textbf {\bibinfo {volume} {12}},\ \bibinfo {pages} {277}
  (\bibinfo {year} {1968})}\BibitemShut {NoStop}%
\bibitem [{\citenamefont {McCumber}(1968)}]{McCumber1968}%
  \BibitemOpen
  \bibfield  {author} {\bibinfo {author} {\bibfnamefont {D.~E.}\ \bibnamefont
  {McCumber}},\ }\href@noop {} {\bibfield  {journal} {\bibinfo  {journal}
  {Journal of Applied Physics}\ }\textbf {\bibinfo {volume} {39}},\ \bibinfo
  {pages} {3113} (\bibinfo {year} {1968})}\BibitemShut {NoStop}%
\bibitem [{\citenamefont {Gregers-Hansen}\ \emph {et~al.}(1971)\citenamefont
  {Gregers-Hansen}, \citenamefont {Levinsen},\ and\ \citenamefont
  {Pedersen}}]{GregersHansenSSC1971}%
  \BibitemOpen
  \bibfield  {author} {\bibinfo {author} {\bibfnamefont {P.~E.}\ \bibnamefont
  {Gregers-Hansen}}, \bibinfo {author} {\bibfnamefont {M.~T.}\ \bibnamefont
  {Levinsen}}, \ and\ \bibinfo {author} {\bibfnamefont {C.~J.}\ \bibnamefont
  {Pedersen}, \bibfnamefont {L.and~Sj{\o}str{\o}m}},\ }\href@noop {} {\bibfield
   {journal} {\bibinfo  {journal} {Solid State Communications}\ }\textbf
  {\bibinfo {volume} {9}},\ \bibinfo {pages} {661} (\bibinfo {year}
  {1971})}\BibitemShut {NoStop}%
\bibitem [{\citenamefont {Gregers-Hansen}\ and\ \citenamefont
  {Levinsen}(1971)}]{GregersHansenLevinsenPRL1971}%
  \BibitemOpen
  \bibfield  {author} {\bibinfo {author} {\bibfnamefont {P.~E.}\ \bibnamefont
  {Gregers-Hansen}}\ and\ \bibinfo {author} {\bibfnamefont {M.~E.}\
  \bibnamefont {Levinsen}},\ }\href@noop {} {\bibfield  {journal} {\bibinfo
  {journal} {Phys. Rev. Lett.}\ }\textbf {\bibinfo {volume} {27}},\ \bibinfo
  {pages} {847} (\bibinfo {year} {1971})}\BibitemShut {NoStop}%
\bibitem [{\citenamefont {Christiansen}\ \emph {et~al.}(1971)\citenamefont
  {Christiansen}, \citenamefont {Hansen},\ and\ \citenamefont
  {Sj{\"o}str{\"o}m}}]{ChristiansenJLTP1971}%
  \BibitemOpen
  \bibfield  {author} {\bibinfo {author} {\bibfnamefont {P.}~\bibnamefont
  {Christiansen}}, \bibinfo {author} {\bibfnamefont {E.}~\bibnamefont
  {Hansen}}, \ and\ \bibinfo {author} {\bibfnamefont {C.~J.}\ \bibnamefont
  {Sj{\"o}str{\"o}m}},\ }\href@noop {} {\bibfield  {journal} {\bibinfo
  {journal} {J. of Low Temp Phys}\ }\textbf {\bibinfo {volume} {4}},\ \bibinfo
  {pages} {349} (\bibinfo {year} {1971})}\BibitemShut {NoStop}%
\bibitem [{\citenamefont {Skocpol}\ \emph
  {et~al.}(1974{\natexlab{a}})\citenamefont {Skocpol}, \citenamefont
  {Beasley},\ and\ \citenamefont {Tinkham}}]{SkocpolJLTP1974}%
  \BibitemOpen
  \bibfield  {author} {\bibinfo {author} {\bibfnamefont {W.~J.}\ \bibnamefont
  {Skocpol}}, \bibinfo {author} {\bibfnamefont {M.~R.}\ \bibnamefont
  {Beasley}}, \ and\ \bibinfo {author} {\bibfnamefont {M.}~\bibnamefont
  {Tinkham}},\ }\href@noop {} {\bibfield  {journal} {\bibinfo  {journal} {J. of
  Low Temp Phys}\ }\textbf {\bibinfo {volume} {16}},\ \bibinfo {pages} {145}
  (\bibinfo {year} {1974}{\natexlab{a}})}\BibitemShut {NoStop}%
\bibitem [{\citenamefont {Mooij}\ \emph {et~al.}(1974)\citenamefont {Mooij},
  \citenamefont {Gorter},\ and\ \citenamefont {Noordam}}]{Mooij1974}%
  \BibitemOpen
  \bibfield  {author} {\bibinfo {author} {\bibfnamefont {J.~E.}\ \bibnamefont
  {Mooij}}, \bibinfo {author} {\bibfnamefont {C.~A.}\ \bibnamefont {Gorter}}, \
  and\ \bibinfo {author} {\bibfnamefont {J.~E.}\ \bibnamefont {Noordam}},\
  }\href@noop {} {\bibfield  {journal} {\bibinfo  {journal} {Revue de Physique
  Appliqu{\'e}e}\ }\textbf {\bibinfo {volume} {9}},\ \bibinfo {pages} {173}
  (\bibinfo {year} {1974})}\BibitemShut {NoStop}%
\bibitem [{\citenamefont {Gubankov}\ \emph {et~al.}(1973)\citenamefont
  {Gubankov}, \citenamefont {Koshelets}, \citenamefont {Likharev},\ and\
  \citenamefont {Ovsyannikov}}]{GubankovVTB1973}%
  \BibitemOpen
  \bibfield  {author} {\bibinfo {author} {\bibfnamefont {V.~N.}\ \bibnamefont
  {Gubankov}}, \bibinfo {author} {\bibfnamefont {V.~P.}\ \bibnamefont
  {Koshelets}}, \bibinfo {author} {\bibfnamefont {K.~K.}\ \bibnamefont
  {Likharev}}, \ and\ \bibinfo {author} {\bibfnamefont {G.~A.}\ \bibnamefont
  {Ovsyannikov}},\ }\href@noop {} {\bibfield  {journal} {\bibinfo  {journal}
  {JETP Letters}\ } (\bibinfo {year} {1973})}\BibitemShut {NoStop}%
\bibitem [{\citenamefont {Gubankov}\ \emph
  {et~al.}(1977{\natexlab{a}})\citenamefont {Gubankov}, \citenamefont
  {Koshelets},\ and\ \citenamefont {Ovsyannikov}}]{Gubankov1976}%
  \BibitemOpen
  \bibfield  {author} {\bibinfo {author} {\bibfnamefont {V.~N.}\ \bibnamefont
  {Gubankov}}, \bibinfo {author} {\bibfnamefont {V.~P.}\ \bibnamefont
  {Koshelets}}, \ and\ \bibinfo {author} {\bibfnamefont {G.~A.}\ \bibnamefont
  {Ovsyannikov}},\ }\href@noop {} {\bibfield  {journal} {\bibinfo  {journal}
  {IEEE Transactions on Magnetics}\ }\textbf {\bibinfo {volume} {MAG-13}},\
  \bibinfo {pages} {228} (\bibinfo {year} {1977}{\natexlab{a}})}\BibitemShut
  {NoStop}%
\bibitem [{\citenamefont {Palmer}\ and\ \citenamefont
  {Decker}(1973)}]{Palmer1973}%
  \BibitemOpen
  \bibfield  {author} {\bibinfo {author} {\bibfnamefont {D.~W.}\ \bibnamefont
  {Palmer}}\ and\ \bibinfo {author} {\bibfnamefont {S.~K.}\ \bibnamefont
  {Decker}},\ }\href@noop {} {\bibfield  {journal} {\bibinfo  {journal} {Review
  of Scientific Instruments}\ }\textbf {\bibinfo {volume} {44}},\ \bibinfo
  {pages} {1621} (\bibinfo {year} {1973})}\BibitemShut {NoStop}%
\bibitem [{\citenamefont {Laibowitz}(1973)}]{Laibowitz1973}%
  \BibitemOpen
  \bibfield  {author} {\bibinfo {author} {\bibfnamefont {R.~B.}\ \bibnamefont
  {Laibowitz}},\ }\href@noop {} {\bibfield  {journal} {\bibinfo  {journal}
  {Appl. Phys. Lett.}\ }\textbf {\bibinfo {volume} {23}},\ \bibinfo {pages}
  {407} (\bibinfo {year} {1973})}\BibitemShut {NoStop}%
\bibitem [{\citenamefont {Skocpol}\ \emph
  {et~al.}(1974{\natexlab{b}})\citenamefont {Skocpol}, \citenamefont
  {Beasley},\ and\ \citenamefont {Tinkham}}]{SBT1974}%
  \BibitemOpen
  \bibfield  {author} {\bibinfo {author} {\bibfnamefont {W.~J.}\ \bibnamefont
  {Skocpol}}, \bibinfo {author} {\bibfnamefont {M.~R.}\ \bibnamefont
  {Beasley}}, \ and\ \bibinfo {author} {\bibfnamefont {M.}~\bibnamefont
  {Tinkham}},\ }\href@noop {} {\bibfield  {journal} {\bibinfo  {journal}
  {Journal of Applied Physics}\ }\textbf {\bibinfo {volume} {45}},\ \bibinfo
  {pages} {4054} (\bibinfo {year} {1974}{\natexlab{b}})}\BibitemShut {NoStop}%
\bibitem [{\citenamefont {Divin}\ \emph {et~al.}(1974)\citenamefont {Divin},
  \citenamefont {Nad'}, \citenamefont {Polyanskii},\ and\ \citenamefont
  {Volkov}}]{Divin1974}%
  \BibitemOpen
  \bibfield  {author} {\bibinfo {author} {\bibfnamefont {Y.~Y.}\ \bibnamefont
  {Divin}}, \bibinfo {author} {\bibfnamefont {F.~Y.}\ \bibnamefont {Nad'}},
  \bibinfo {author} {\bibfnamefont {O.~Y.}\ \bibnamefont {Polyanskii}}, \ and\
  \bibinfo {author} {\bibfnamefont {A.~F.}\ \bibnamefont {Volkov}},\
  }\href@noop {} {\bibfield  {journal} {\bibinfo  {journal} {Revue de Physique
  Appliqu{\'e}e}\ }\textbf {\bibinfo {volume} {9}},\ \bibinfo {pages} {125}
  (\bibinfo {year} {1974})}\BibitemShut {NoStop}%
\bibitem [{\citenamefont {Kulik}\ and\ \citenamefont
  {Omel'yanchuk}(1975)}]{KOI1975}%
  \BibitemOpen
  \bibfield  {author} {\bibinfo {author} {\bibfnamefont {I.~O.}\ \bibnamefont
  {Kulik}}\ and\ \bibinfo {author} {\bibfnamefont {A.~N.}\ \bibnamefont
  {Omel'yanchuk}},\ }\href@noop {} {\bibfield  {journal} {\bibinfo  {journal}
  {JETP Lett.}\ }\textbf {\bibinfo {volume} {21}},\ \bibinfo {pages} {96}
  (\bibinfo {year} {1975})}\BibitemShut {NoStop}%
\bibitem [{\citenamefont {Klapwijk}\ and\ \citenamefont
  {Veenstra}(1974)}]{KlapwijkVeenstra1974}%
  \BibitemOpen
  \bibfield  {author} {\bibinfo {author} {\bibfnamefont {T.~M.}\ \bibnamefont
  {Klapwijk}}\ and\ \bibinfo {author} {\bibfnamefont {T.~B.}\ \bibnamefont
  {Veenstra}},\ }\href@noop {} {\bibfield  {journal} {\bibinfo  {journal}
  {Physics Letters A}\ }\textbf {\bibinfo {volume} {47}},\ \bibinfo {pages}
  {351} (\bibinfo {year} {1974})}\BibitemShut {NoStop}%
\bibitem [{\citenamefont {Klapwijk}\ and\ \citenamefont
  {Mooij}(1975)}]{KlapwijkMooij1975}%
  \BibitemOpen
  \bibfield  {author} {\bibinfo {author} {\bibfnamefont {T.~M.}\ \bibnamefont
  {Klapwijk}}\ and\ \bibinfo {author} {\bibfnamefont {J.~E.}\ \bibnamefont
  {Mooij}},\ }\href@noop {} {\bibfield  {journal} {\bibinfo  {journal} {IEEE
  Transactions on Magnetics}\ }\textbf {\bibinfo {volume} {MAG-11}},\ \bibinfo
  {pages} {858} (\bibinfo {year} {1975})}\BibitemShut {NoStop}%
\bibitem [{\citenamefont {Octavio}\ \emph {et~al.}(1977)\citenamefont
  {Octavio}, \citenamefont {Skocpol},\ and\ \citenamefont
  {Tinkham}}]{Octavio1977}%
  \BibitemOpen
  \bibfield  {author} {\bibinfo {author} {\bibfnamefont {M.}~\bibnamefont
  {Octavio}}, \bibinfo {author} {\bibfnamefont {W.~J.}\ \bibnamefont
  {Skocpol}}, \ and\ \bibinfo {author} {\bibfnamefont {M.}~\bibnamefont
  {Tinkham}},\ }\href@noop {} {\bibfield  {journal} {\bibinfo  {journal} {IEEE
  Transactions on Magnetics}\ }\textbf {\bibinfo {volume} {MAG-13}},\ \bibinfo
  {pages} {739} (\bibinfo {year} {1977})}\BibitemShut {NoStop}%
\bibitem [{\citenamefont {Gubankov}\ \emph
  {et~al.}(1977{\natexlab{b}})\citenamefont {Gubankov}, \citenamefont
  {Koshelets},\ and\ \citenamefont {Ovsyannikov}}]{GubankovVTB1977}%
  \BibitemOpen
  \bibfield  {author} {\bibinfo {author} {\bibfnamefont {V.~N.}\ \bibnamefont
  {Gubankov}}, \bibinfo {author} {\bibfnamefont {V.~P.}\ \bibnamefont
  {Koshelets}}, \ and\ \bibinfo {author} {\bibfnamefont {G.~A.}\ \bibnamefont
  {Ovsyannikov}},\ }\href@noop {} {\bibfield  {journal} {\bibinfo  {journal}
  {Soviet Physics JETP}\ }\textbf {\bibinfo {volume} {46}},\ \bibinfo {pages}
  {755} (\bibinfo {year} {1977}{\natexlab{b}})}\BibitemShut {NoStop}%
\bibitem [{\citenamefont {van Ruitenbeek}\ \emph {et~al.}(1996)\citenamefont
  {van Ruitenbeek}, \citenamefont {Alvarez}, \citenamefont {Pi{\~n}eyro},
  \citenamefont {Grahmann}, \citenamefont {Joyez}, \citenamefont {Devoret},
  \citenamefont {Esteve},\ and\ \citenamefont {Urbina}}]{Ruitenbeek1996}%
  \BibitemOpen
  \bibfield  {author} {\bibinfo {author} {\bibfnamefont {J.~M.}\ \bibnamefont
  {van Ruitenbeek}}, \bibinfo {author} {\bibfnamefont {A.}~\bibnamefont
  {Alvarez}}, \bibinfo {author} {\bibfnamefont {I.}~\bibnamefont
  {Pi{\~n}eyro}}, \bibinfo {author} {\bibfnamefont {C.}~\bibnamefont
  {Grahmann}}, \bibinfo {author} {\bibfnamefont {P.}~\bibnamefont {Joyez}},
  \bibinfo {author} {\bibfnamefont {M.~H.}\ \bibnamefont {Devoret}}, \bibinfo
  {author} {\bibfnamefont {D.}~\bibnamefont {Esteve}}, \ and\ \bibinfo {author}
  {\bibfnamefont {C.}~\bibnamefont {Urbina}},\ }\href@noop {} {\bibfield
  {journal} {\bibinfo  {journal} {Review of Scientific Instruments}\ }\textbf
  {\bibinfo {volume} {67}},\ \bibinfo {pages} {108} (\bibinfo {year}
  {1996})}\BibitemShut {NoStop}%
\bibitem [{\citenamefont {Notarys}\ \emph {et~al.}(1973)\citenamefont
  {Notarys}, \citenamefont {Yu},\ and\ \citenamefont
  {Mercereau}}]{NotarysPRL1973}%
  \BibitemOpen
  \bibfield  {author} {\bibinfo {author} {\bibfnamefont {H.~A.}\ \bibnamefont
  {Notarys}}, \bibinfo {author} {\bibfnamefont {M.~L.}\ \bibnamefont {Yu}}, \
  and\ \bibinfo {author} {\bibfnamefont {J.~E.}\ \bibnamefont {Mercereau}},\
  }\href@noop {} {\bibfield  {journal} {\bibinfo  {journal} {Phys. Rev. Lett.}\
  }\textbf {\bibinfo {volume} {30}},\ \bibinfo {pages} {743} (\bibinfo {year}
  {1973})}\BibitemShut {NoStop}%
\bibitem [{\citenamefont {Warlaumont}\ \emph {et~al.}(1979)\citenamefont
  {Warlaumont}, \citenamefont {Brown}, \citenamefont {Foxe},\ and\
  \citenamefont {Buhrman}}]{WarlaumontPRL1979}%
  \BibitemOpen
  \bibfield  {author} {\bibinfo {author} {\bibfnamefont {J.~M.}\ \bibnamefont
  {Warlaumont}}, \bibinfo {author} {\bibfnamefont {J.~C.}\ \bibnamefont
  {Brown}}, \bibinfo {author} {\bibfnamefont {T.}~\bibnamefont {Foxe}}, \ and\
  \bibinfo {author} {\bibfnamefont {R.~A.}\ \bibnamefont {Buhrman}},\
  }\href@noop {} {\bibfield  {journal} {\bibinfo  {journal} {Phys. Rev. Lett.}\
  }\textbf {\bibinfo {volume} {43}},\ \bibinfo {pages} {169} (\bibinfo {year}
  {1979})}\BibitemShut {NoStop}%
\bibitem [{\citenamefont {de~Visser}\ \emph {et~al.}(2014)\citenamefont
  {de~Visser}, \citenamefont {Goldie}, \citenamefont {Diener}, \citenamefont
  {Withington}, \citenamefont {Baselmans},\ and\ \citenamefont
  {Klapwijk}}]{DeVisserPRL2014}%
  \BibitemOpen
  \bibfield  {author} {\bibinfo {author} {\bibfnamefont {P.~J.}\ \bibnamefont
  {de~Visser}}, \bibinfo {author} {\bibfnamefont {D.~J.}\ \bibnamefont
  {Goldie}}, \bibinfo {author} {\bibfnamefont {P.}~\bibnamefont {Diener}},
  \bibinfo {author} {\bibfnamefont {S.}~\bibnamefont {Withington}}, \bibinfo
  {author} {\bibfnamefont {J.~J.~A.}\ \bibnamefont {Baselmans}}, \ and\
  \bibinfo {author} {\bibfnamefont {T.~M.}\ \bibnamefont {Klapwijk}},\
  }\href@noop {} {\bibfield  {journal} {\bibinfo  {journal} {Phys. Rev. Lett.}\
  }\textbf {\bibinfo {volume} {112}},\ \bibinfo {pages} {047004} (\bibinfo
  {year} {2014})}\BibitemShut {NoStop}%
\bibitem [{\citenamefont {Parmenter}(1961)}]{Parmenter1961}%
  \BibitemOpen
  \bibfield  {author} {\bibinfo {author} {\bibfnamefont {R.~H.}\ \bibnamefont
  {Parmenter}},\ }\href@noop {} {\bibfield  {journal} {\bibinfo  {journal}
  {Phys. Rev. Lett.}\ }\textbf {\bibinfo {volume} {7}},\ \bibinfo {pages} {274}
  (\bibinfo {year} {1961})}\BibitemShut {NoStop}%
\bibitem [{\citenamefont {Blamire}\ \emph {et~al.}(1991)\citenamefont
  {Blamire}, \citenamefont {Kirk}, \citenamefont {Evetts},\ and\ \citenamefont
  {Klapwijk}}]{BlamirePRL1991}%
  \BibitemOpen
  \bibfield  {author} {\bibinfo {author} {\bibfnamefont {M.~G.}\ \bibnamefont
  {Blamire}}, \bibinfo {author} {\bibfnamefont {E.~C.~G.}\ \bibnamefont
  {Kirk}}, \bibinfo {author} {\bibfnamefont {J.~E.}\ \bibnamefont {Evetts}}, \
  and\ \bibinfo {author} {\bibfnamefont {T.~M.}\ \bibnamefont {Klapwijk}},\
  }\href@noop {} {\bibfield  {journal} {\bibinfo  {journal} {Phys. Rev. Lett.}\
  }\textbf {\bibinfo {volume} {66}},\ \bibinfo {pages} {220} (\bibinfo {year}
  {1991})}\BibitemShut {NoStop}%
\bibitem [{\citenamefont {Chang}\ and\ \citenamefont
  {Scalapino}(1977)}]{ChangScalapino1977}%
  \BibitemOpen
  \bibfield  {author} {\bibinfo {author} {\bibfnamefont {J.~J.}\ \bibnamefont
  {Chang}}\ and\ \bibinfo {author} {\bibfnamefont {D.~J.}\ \bibnamefont
  {Scalapino}},\ }\href@noop {} {\bibfield  {journal} {\bibinfo  {journal} {J.
  of Low Temp Phys}\ }\textbf {\bibinfo {volume} {29}},\ \bibinfo {pages} {477}
  (\bibinfo {year} {1977})}\BibitemShut {NoStop}%
\bibitem [{\citenamefont {Klapwijk}\ and\ \citenamefont
  {Mooij}(1976)}]{KlapwijkMooij1976}%
  \BibitemOpen
  \bibfield  {author} {\bibinfo {author} {\bibfnamefont {T.~M.}\ \bibnamefont
  {Klapwijk}}\ and\ \bibinfo {author} {\bibfnamefont {J.~E.}\ \bibnamefont
  {Mooij}},\ }\href@noop {} {\bibfield  {journal} {\bibinfo  {journal} {Physica
  B}\ }\textbf {\bibinfo {volume} {81}},\ \bibinfo {pages} {132} (\bibinfo
  {year} {1976})}\BibitemShut {NoStop}%
\bibitem [{\citenamefont {Schmid}(1977)}]{Schmid1977}%
  \BibitemOpen
  \bibfield  {author} {\bibinfo {author} {\bibfnamefont {A.}~\bibnamefont
  {Schmid}},\ }\href@noop {} {\bibfield  {journal} {\bibinfo  {journal} {Phys.
  Rev. Lett.}\ }\textbf {\bibinfo {volume} {38}},\ \bibinfo {pages} {922}
  (\bibinfo {year} {1977})}\BibitemShut {NoStop}%
\bibitem [{\citenamefont {Latyshev}\ and\ \citenamefont
  {Nad'}(1974)}]{Latyshev1974}%
  \BibitemOpen
  \bibfield  {author} {\bibinfo {author} {\bibfnamefont {Y.~I.}\ \bibnamefont
  {Latyshev}}\ and\ \bibinfo {author} {\bibfnamefont {F.~Y.}\ \bibnamefont
  {Nad'}},\ }\href@noop {} {\bibfield  {journal} {\bibinfo  {journal} {JETP
  Letters}\ }\textbf {\bibinfo {volume} {19}},\ \bibinfo {pages} {380}
  (\bibinfo {year} {1974})}\BibitemShut {NoStop}%
\bibitem [{\citenamefont {Latyshev}\ and\ \citenamefont
  {Nad'}(1976)}]{Latyshev1976}%
  \BibitemOpen
  \bibfield  {author} {\bibinfo {author} {\bibfnamefont {Y.~I.}\ \bibnamefont
  {Latyshev}}\ and\ \bibinfo {author} {\bibfnamefont {F.~Y.}\ \bibnamefont
  {Nad'}},\ }\href@noop {} {\bibfield  {journal} {\bibinfo  {journal} {Soviet
  Physics JETP}\ }\textbf {\bibinfo {volume} {44}},\ \bibinfo {pages} {1136}
  (\bibinfo {year} {1976})}\BibitemShut {NoStop}%
\bibitem [{\citenamefont {Tredwell}\ and\ \citenamefont
  {Jacobsen}(1975)}]{TredwellPRL1975}%
  \BibitemOpen
  \bibfield  {author} {\bibinfo {author} {\bibfnamefont {T.~J.}\ \bibnamefont
  {Tredwell}}\ and\ \bibinfo {author} {\bibfnamefont {E.~H.}\ \bibnamefont
  {Jacobsen}},\ }\href@noop {} {\bibfield  {journal} {\bibinfo  {journal}
  {Phys. Rev. Lett.}\ }\textbf {\bibinfo {volume} {35}},\ \bibinfo {pages}
  {244} (\bibinfo {year} {1975})}\BibitemShut {NoStop}%
\bibitem [{\citenamefont {Tredwell}\ and\ \citenamefont
  {Jacobsen}(1976)}]{TredwellPRB1976}%
  \BibitemOpen
  \bibfield  {author} {\bibinfo {author} {\bibfnamefont {T.~J.}\ \bibnamefont
  {Tredwell}}\ and\ \bibinfo {author} {\bibfnamefont {E.~H.}\ \bibnamefont
  {Jacobsen}},\ }\href@noop {} {\bibfield  {journal} {\bibinfo  {journal}
  {Phys. Rev. B}\ }\textbf {\bibinfo {volume} {13}},\ \bibinfo {pages} {2931}
  (\bibinfo {year} {1976})}\BibitemShut {NoStop}%
\bibitem [{\citenamefont {Kommers}\ and\ \citenamefont
  {Clarke}(1977)}]{Kommers1977}%
  \BibitemOpen
  \bibfield  {author} {\bibinfo {author} {\bibfnamefont {T.}~\bibnamefont
  {Kommers}}\ and\ \bibinfo {author} {\bibfnamefont {J.}~\bibnamefont
  {Clarke}},\ }\href@noop {} {\bibfield  {journal} {\bibinfo  {journal} {Phys.
  Rev. Lett.}\ }\textbf {\bibinfo {volume} {38}},\ \bibinfo {pages} {1091}
  (\bibinfo {year} {1977})}\BibitemShut {NoStop}%
\bibitem [{\citenamefont {Dahlberg}\ \emph {et~al.}(1979)\citenamefont
  {Dahlberg}, \citenamefont {Orbach},\ and\ \citenamefont
  {Schuller}}]{DahlbergOrbachSchuller1979}%
  \BibitemOpen
  \bibfield  {author} {\bibinfo {author} {\bibfnamefont {E.~D.}\ \bibnamefont
  {Dahlberg}}, \bibinfo {author} {\bibfnamefont {R.~L.}\ \bibnamefont
  {Orbach}}, \ and\ \bibinfo {author} {\bibfnamefont {I.}~\bibnamefont
  {Schuller}},\ }\href@noop {} {\bibfield  {journal} {\bibinfo  {journal} {J.
  of Low Temp Phys}\ }\textbf {\bibinfo {volume} {36}},\ \bibinfo {pages} {367}
  (\bibinfo {year} {1979})}\BibitemShut {NoStop}%
\bibitem [{\citenamefont {Hall}\ \emph {et~al.}(1980)\citenamefont {Hall},
  \citenamefont {Holdeman},\ and\ \citenamefont {Soulen~Jr.}}]{HallSoulen1980}%
  \BibitemOpen
  \bibfield  {author} {\bibinfo {author} {\bibfnamefont {J.~T.}\ \bibnamefont
  {Hall}}, \bibinfo {author} {\bibfnamefont {L.~B.}\ \bibnamefont {Holdeman}},
  \ and\ \bibinfo {author} {\bibfnamefont {R.~J.}\ \bibnamefont {Soulen~Jr.}},\
  }\href@noop {} {\bibfield  {journal} {\bibinfo  {journal} {Phys. Rev. Lett.}\
  }\textbf {\bibinfo {volume} {45}},\ \bibinfo {pages} {1011} (\bibinfo {year}
  {1980})}\BibitemShut {NoStop}%
\bibitem [{\citenamefont {Bardeen}(1962)}]{Bardeen1962}%
  \BibitemOpen
  \bibfield  {author} {\bibinfo {author} {\bibfnamefont {J.}~\bibnamefont
  {Bardeen}},\ }\href@noop {} {\bibfield  {journal} {\bibinfo  {journal} {Rev.
  Mod. Phys.}\ }\textbf {\bibinfo {volume} {34}},\ \bibinfo {pages} {667}
  (\bibinfo {year} {1962})}\BibitemShut {NoStop}%
\bibitem [{\citenamefont {Anthore}\ \emph {et~al.}(2003)\citenamefont
  {Anthore}, \citenamefont {Pothier},\ and\ \citenamefont
  {Esteve}}]{Anthore2003}%
  \BibitemOpen
  \bibfield  {author} {\bibinfo {author} {\bibfnamefont {A.}~\bibnamefont
  {Anthore}}, \bibinfo {author} {\bibfnamefont {H.}~\bibnamefont {Pothier}}, \
  and\ \bibinfo {author} {\bibfnamefont {D.}~\bibnamefont {Esteve}},\
  }\href@noop {} {\bibfield  {journal} {\bibinfo  {journal} {Phys. Rev. Lett.}\
  }\textbf {\bibinfo {volume} {90}},\ \bibinfo {pages} {127001} (\bibinfo
  {year} {2003})}\BibitemShut {NoStop}%
\bibitem [{\citenamefont {Usadel}(1970)}]{Usadel1970}%
  \BibitemOpen
  \bibfield  {author} {\bibinfo {author} {\bibfnamefont {K.~D.}\ \bibnamefont
  {Usadel}},\ }\href@noop {} {\bibfield  {journal} {\bibinfo  {journal} {Phys.
  Rev. Lett.}\ }\textbf {\bibinfo {volume} {25}},\ \bibinfo {pages} {507}
  (\bibinfo {year} {1970})}\BibitemShut {NoStop}%
\bibitem [{\citenamefont {Romijn}\ \emph {et~al.}(1982)\citenamefont {Romijn},
  \citenamefont {Klapwijk}, \citenamefont {Renne},\ and\ \citenamefont
  {Mooij}}]{Romijn1982}%
  \BibitemOpen
  \bibfield  {author} {\bibinfo {author} {\bibfnamefont {J.}~\bibnamefont
  {Romijn}}, \bibinfo {author} {\bibfnamefont {T.~M.}\ \bibnamefont
  {Klapwijk}}, \bibinfo {author} {\bibfnamefont {M.}~\bibnamefont {Renne}}, \
  and\ \bibinfo {author} {\bibfnamefont {J.}~\bibnamefont {Mooij}},\
  }\href@noop {} {\bibfield  {journal} {\bibinfo  {journal} {Phys. Rev. B}\
  }\textbf {\bibinfo {volume} {26}},\ \bibinfo {pages} {3648} (\bibinfo {year}
  {1982})}\BibitemShut {NoStop}%
\bibitem [{\citenamefont {Falco}\ \emph {et~al.}(1980)\citenamefont {Falco},
  \citenamefont {Werner},\ and\ \citenamefont {Schuller}}]{FalcoSSC1980}%
  \BibitemOpen
  \bibfield  {author} {\bibinfo {author} {\bibfnamefont {C.~M.}\ \bibnamefont
  {Falco}}, \bibinfo {author} {\bibfnamefont {T.~R.}\ \bibnamefont {Werner}}, \
  and\ \bibinfo {author} {\bibfnamefont {I.~K.}\ \bibnamefont {Schuller}},\
  }\href@noop {} {\bibfield  {journal} {\bibinfo  {journal} {Solid State
  Communications}\ }\textbf {\bibinfo {volume} {34}},\ \bibinfo {pages} {535}
  (\bibinfo {year} {1980})}\BibitemShut {NoStop}%
\bibitem [{\citenamefont {Mooij}\ \emph {et~al.}(1980)\citenamefont {Mooij},
  \citenamefont {Lambert},\ and\ \citenamefont {Klapwijk}}]{MooijSSC1980}%
  \BibitemOpen
  \bibfield  {author} {\bibinfo {author} {\bibfnamefont {J.~E.}\ \bibnamefont
  {Mooij}}, \bibinfo {author} {\bibfnamefont {N.}~\bibnamefont {Lambert}}, \
  and\ \bibinfo {author} {\bibfnamefont {T.~M.}\ \bibnamefont {Klapwijk}},\
  }\href@noop {} {\bibfield  {journal} {\bibinfo  {journal} {Solid State
  Communications}\ }\textbf {\bibinfo {volume} {36}},\ \bibinfo {pages} {585}
  (\bibinfo {year} {1980})}\BibitemShut {NoStop}%
\bibitem [{\citenamefont {Notarys}\ and\ \citenamefont
  {Mercereau}(1973)}]{NotarysJAP1973}%
  \BibitemOpen
  \bibfield  {author} {\bibinfo {author} {\bibfnamefont {H.~A.}\ \bibnamefont
  {Notarys}}\ and\ \bibinfo {author} {\bibfnamefont {J.~E.}\ \bibnamefont
  {Mercereau}},\ }\href@noop {} {\bibfield  {journal} {\bibinfo  {journal}
  {Journal of Applied Physics}\ }\textbf {\bibinfo {volume} {44}},\ \bibinfo
  {pages} {1821} (\bibinfo {year} {1973})}\BibitemShut {NoStop}%
\bibitem [{\citenamefont {Tikhonov}\ \emph {et~al.}(2018)\citenamefont
  {Tikhonov}, \citenamefont {Skvortsov},\ and\ \citenamefont
  {Klapwijk}}]{TikhonovPRB2018}%
  \BibitemOpen
  \bibfield  {author} {\bibinfo {author} {\bibfnamefont {K.~S.}\ \bibnamefont
  {Tikhonov}}, \bibinfo {author} {\bibfnamefont {M.~A.}\ \bibnamefont
  {Skvortsov}}, \ and\ \bibinfo {author} {\bibfnamefont {T.~M.}\ \bibnamefont
  {Klapwijk}},\ }\href@noop {} {\bibfield  {journal} {\bibinfo  {journal}
  {Phys. Rev. B}\ }\textbf {\bibinfo {volume} {97}},\ \bibinfo {pages} {184516}
  (\bibinfo {year} {2018})}\BibitemShut {NoStop}%
\bibitem [{\citenamefont {Deutscher}\ and\ \citenamefont
  {De~Gennes}(1969)}]{ParksII1969}%
  \BibitemOpen
  \bibfield  {author} {\bibinfo {author} {\bibfnamefont {G.}~\bibnamefont
  {Deutscher}}\ and\ \bibinfo {author} {\bibfnamefont {P.~G.}\ \bibnamefont
  {De~Gennes}},\ }in\ \href@noop {} {\emph {\bibinfo {booktitle}
  {Superconductivity}}},\ Vol.~\bibinfo {volume} {II},\ \bibinfo {editor}
  {edited by\ \bibinfo {editor} {\bibfnamefont {R.~D.}\ \bibnamefont {Parks}}}\
  (\bibinfo  {publisher} {Marcel Dekker Inc,},\ \bibinfo {address} {New York},\
  \bibinfo {year} {1969})\ Chap.~\bibinfo {chapter} {17}, pp.\ \bibinfo {pages}
  {1005--1034}\BibitemShut {NoStop}%
\bibitem [{\citenamefont {Volkov}\ \emph {et~al.}(1993)\citenamefont {Volkov},
  \citenamefont {Zaitsev},\ and\ \citenamefont {Klapwijk}}]{VZK1993}%
  \BibitemOpen
  \bibfield  {author} {\bibinfo {author} {\bibfnamefont {A.~F.}\ \bibnamefont
  {Volkov}}, \bibinfo {author} {\bibfnamefont {A.~V.}\ \bibnamefont {Zaitsev}},
  \ and\ \bibinfo {author} {\bibfnamefont {T.~M.}\ \bibnamefont {Klapwijk}},\
  }\href@noop {} {\bibfield  {journal} {\bibinfo  {journal} {Physica C}\
  }\textbf {\bibinfo {volume} {210}},\ \bibinfo {pages} {21} (\bibinfo {year}
  {1993})}\BibitemShut {NoStop}%
\bibitem [{\citenamefont {Belzig}\ \emph {et~al.}(1999)\citenamefont {Belzig},
  \citenamefont {Wilhelm}, \citenamefont {Bruder},\ and\ \citenamefont
  {Sch{\"o}n}}]{Belzig1999}%
  \BibitemOpen
  \bibfield  {author} {\bibinfo {author} {\bibfnamefont {W.}~\bibnamefont
  {Belzig}}, \bibinfo {author} {\bibfnamefont {F.~K.}\ \bibnamefont {Wilhelm}},
  \bibinfo {author} {\bibfnamefont {C.}~\bibnamefont {Bruder}}, \ and\ \bibinfo
  {author} {\bibfnamefont {G.}~\bibnamefont {Sch{\"o}n}},\ }\href@noop {}
  {\bibfield  {journal} {\bibinfo  {journal} {Superlattices and
  microstructures}\ }\textbf {\bibinfo {volume} {25}},\ \bibinfo {pages} {1251}
  (\bibinfo {year} {1999})}\BibitemShut {NoStop}%
\bibitem [{\citenamefont {Klapwijk}(2004)}]{Klapwijk2004}%
  \BibitemOpen
  \bibfield  {author} {\bibinfo {author} {\bibfnamefont {T.~M.}\ \bibnamefont
  {Klapwijk}},\ }\href@noop {} {\bibfield  {journal} {\bibinfo  {journal}
  {Journal of Superconductivity}\ }\textbf {\bibinfo {volume} {17}},\ \bibinfo
  {pages} {593} (\bibinfo {year} {2004})}\BibitemShut {NoStop}%
\bibitem [{\citenamefont {Baselmans}\ \emph {et~al.}(2001)\citenamefont
  {Baselmans}, \citenamefont {van Wees},\ and\ \citenamefont
  {Klapwijk}}]{BaselmansPRB2001}%
  \BibitemOpen
  \bibfield  {author} {\bibinfo {author} {\bibfnamefont {J.~J.~A.}\
  \bibnamefont {Baselmans}}, \bibinfo {author} {\bibfnamefont {B.~J.}\
  \bibnamefont {van Wees}}, \ and\ \bibinfo {author} {\bibfnamefont {T.~M.}\
  \bibnamefont {Klapwijk}},\ }\href@noop {} {\bibfield  {journal} {\bibinfo
  {journal} {Phys. Rev. B}\ }\textbf {\bibinfo {volume} {63}},\ \bibinfo
  {pages} {094504} (\bibinfo {year} {2001})}\BibitemShut {NoStop}%
\bibitem [{\citenamefont {Morpurgo}\ \emph {et~al.}(1998)\citenamefont
  {Morpurgo}, \citenamefont {Klapwijk},\ and\ \citenamefont {van
  Wees}}]{Morpurgo1998}%
  \BibitemOpen
  \bibfield  {author} {\bibinfo {author} {\bibfnamefont {A.~F.}\ \bibnamefont
  {Morpurgo}}, \bibinfo {author} {\bibfnamefont {T.~M.}\ \bibnamefont
  {Klapwijk}}, \ and\ \bibinfo {author} {\bibfnamefont {B.~J.}\ \bibnamefont
  {van Wees}},\ }\href@noop {} {\bibfield  {journal} {\bibinfo  {journal}
  {Appl. Phys. Lett.}\ }\textbf {\bibinfo {volume} {72}},\ \bibinfo {pages}
  {966} (\bibinfo {year} {1998})}\BibitemShut {NoStop}%
\bibitem [{\citenamefont {Baselmans}\ \emph {et~al.}(1998)\citenamefont
  {Baselmans}, \citenamefont {Morpurgo}, \citenamefont {van Wees},\ and\
  \citenamefont {Klapwijk}}]{BaselmansNature1998}%
  \BibitemOpen
  \bibfield  {author} {\bibinfo {author} {\bibfnamefont {J.~J.~A.}\
  \bibnamefont {Baselmans}}, \bibinfo {author} {\bibfnamefont {A.~F.}\
  \bibnamefont {Morpurgo}}, \bibinfo {author} {\bibfnamefont {B.~J.}\
  \bibnamefont {van Wees}}, \ and\ \bibinfo {author} {\bibfnamefont {T.~M.}\
  \bibnamefont {Klapwijk}},\ }\href@noop {} {\bibfield  {journal} {\bibinfo
  {journal} {Nature}\ }\textbf {\bibinfo {volume} {397}},\ \bibinfo {pages}
  {43} (\bibinfo {year} {1998})}\BibitemShut {NoStop}%
\bibitem [{\citenamefont {Baselmans}\ \emph {et~al.}(2002)\citenamefont
  {Baselmans}, \citenamefont {Heikkil{\"a}}, \citenamefont {van Wees},\ and\
  \citenamefont {Klapwijk}}]{BaselmansPRL2002}%
  \BibitemOpen
  \bibfield  {author} {\bibinfo {author} {\bibfnamefont {J.~J.~A.}\
  \bibnamefont {Baselmans}}, \bibinfo {author} {\bibfnamefont {T.~T.}\
  \bibnamefont {Heikkil{\"a}}}, \bibinfo {author} {\bibfnamefont {B.~J.}\
  \bibnamefont {van Wees}}, \ and\ \bibinfo {author} {\bibfnamefont {T.~M.}\
  \bibnamefont {Klapwijk}},\ }\href@noop {} {\bibfield  {journal} {\bibinfo
  {journal} {Phys. Rev. Lett.}\ }\textbf {\bibinfo {volume} {89}},\ \bibinfo
  {pages} {207002} (\bibinfo {year} {2002})}\BibitemShut {NoStop}%
\bibitem [{\citenamefont {Kozub}\ and\ \citenamefont
  {Rudin}(1995)}]{Kozub1995}%
  \BibitemOpen
  \bibfield  {author} {\bibinfo {author} {\bibfnamefont {V.~I.}\ \bibnamefont
  {Kozub}}\ and\ \bibinfo {author} {\bibfnamefont {A.~M.}\ \bibnamefont
  {Rudin}},\ }\href@noop {} {\bibfield  {journal} {\bibinfo  {journal} {Phys.
  Rev. B}\ }\textbf {\bibinfo {volume} {52}},\ \bibinfo {pages} {7853}
  (\bibinfo {year} {1995})}\BibitemShut {NoStop}%
\bibitem [{\citenamefont {Pothier}\ \emph {et~al.}(1997)\citenamefont
  {Pothier}, \citenamefont {Gu\'eron}, \citenamefont {Birge}, \citenamefont
  {Esteve},\ and\ \citenamefont {Devoret}}]{Pothier1997}%
  \BibitemOpen
  \bibfield  {author} {\bibinfo {author} {\bibfnamefont {H.}~\bibnamefont
  {Pothier}}, \bibinfo {author} {\bibfnamefont {S.}~\bibnamefont {Gu\'eron}},
  \bibinfo {author} {\bibfnamefont {N.~O.}\ \bibnamefont {Birge}}, \bibinfo
  {author} {\bibfnamefont {D.}~\bibnamefont {Esteve}}, \ and\ \bibinfo {author}
  {\bibfnamefont {M.}~\bibnamefont {Devoret}},\ }\href@noop {} {\bibfield
  {journal} {\bibinfo  {journal} {Phys. Rev. Lett.}\ }\textbf {\bibinfo
  {volume} {79}},\ \bibinfo {pages} {3490} (\bibinfo {year}
  {1997})}\BibitemShut {NoStop}%
\bibitem [{\citenamefont {Schmid}\ \emph {et~al.}(1980)\citenamefont {Schmid},
  \citenamefont {Sch{\"o}n},\ and\ \citenamefont {Tinkham}}]{SST1980}%
  \BibitemOpen
  \bibfield  {author} {\bibinfo {author} {\bibfnamefont {A.}~\bibnamefont
  {Schmid}}, \bibinfo {author} {\bibfnamefont {G.}~\bibnamefont {Sch{\"o}n}}, \
  and\ \bibinfo {author} {\bibfnamefont {M.}~\bibnamefont {Tinkham}},\
  }\href@noop {} {\bibfield  {journal} {\bibinfo  {journal} {Phys. Rev. B}\
  }\textbf {\bibinfo {volume} {21}},\ \bibinfo {pages} {5076} (\bibinfo {year}
  {1980})}\BibitemShut {NoStop}%
\bibitem [{\citenamefont {Courtois}\ \emph {et~al.}(2008)\citenamefont
  {Courtois}, \citenamefont {Meschke}, \citenamefont {Peltonen},\ and\
  \citenamefont {Pekola}}]{Courtois2008}%
  \BibitemOpen
  \bibfield  {author} {\bibinfo {author} {\bibfnamefont {H.}~\bibnamefont
  {Courtois}}, \bibinfo {author} {\bibfnamefont {M.}~\bibnamefont {Meschke}},
  \bibinfo {author} {\bibfnamefont {J.~T.}\ \bibnamefont {Peltonen}}, \ and\
  \bibinfo {author} {\bibfnamefont {J.~P.}\ \bibnamefont {Pekola}},\
  }\href@noop {} {\bibfield  {journal} {\bibinfo  {journal} {Phys. Rev. Lett.}\
  }\textbf {\bibinfo {volume} {101}},\ \bibinfo {pages} {067002} (\bibinfo
  {year} {2008})}\BibitemShut {NoStop}%
\bibitem [{\citenamefont {De~Cecco}\ \emph {et~al.}(2016)\citenamefont
  {De~Cecco}, \citenamefont {Le~Calvez}, \citenamefont {Sac{\'e}p{\'e}},
  \citenamefont {Winkelmann},\ and\ \citenamefont
  {Courtois}}]{DeCeccoCourtois2016}%
  \BibitemOpen
  \bibfield  {author} {\bibinfo {author} {\bibfnamefont {A.}~\bibnamefont
  {De~Cecco}}, \bibinfo {author} {\bibfnamefont {K.}~\bibnamefont {Le~Calvez}},
  \bibinfo {author} {\bibfnamefont {B.}~\bibnamefont {Sac{\'e}p{\'e}}},
  \bibinfo {author} {\bibfnamefont {C.~B.}\ \bibnamefont {Winkelmann}}, \ and\
  \bibinfo {author} {\bibfnamefont {H.}~\bibnamefont {Courtois}},\ }\href@noop
  {} {\bibfield  {journal} {\bibinfo  {journal} {Phys. Rev. B}\ }\textbf
  {\bibinfo {volume} {93}},\ \bibinfo {pages} {18505R} (\bibinfo {year}
  {2016})}\BibitemShut {NoStop}%
\bibitem [{\citenamefont {Tinkham}\ \emph {et~al.}(1977)\citenamefont
  {Tinkham}, \citenamefont {Octavio},\ and\ \citenamefont
  {Skocpol}}]{Tinkham1977}%
  \BibitemOpen
  \bibfield  {author} {\bibinfo {author} {\bibfnamefont {M.}~\bibnamefont
  {Tinkham}}, \bibinfo {author} {\bibfnamefont {M.}~\bibnamefont {Octavio}}, \
  and\ \bibinfo {author} {\bibfnamefont {W.}~\bibnamefont {Skocpol}},\
  }\href@noop {} {\bibfield  {journal} {\bibinfo  {journal} {Journal of Applied
  Physics}\ }\textbf {\bibinfo {volume} {48}},\ \bibinfo {pages} {1311}
  (\bibinfo {year} {1977})}\BibitemShut {NoStop}%
\bibitem [{\citenamefont {Virtanen}\ \emph {et~al.}(2010)\citenamefont
  {Virtanen}, \citenamefont {Heikkil{\"a}}, \citenamefont {Bergeret},\ and\
  \citenamefont {Cuevas}}]{Virtanen2010}%
  \BibitemOpen
  \bibfield  {author} {\bibinfo {author} {\bibfnamefont {P.}~\bibnamefont
  {Virtanen}}, \bibinfo {author} {\bibfnamefont {T.~T.}\ \bibnamefont
  {Heikkil{\"a}}}, \bibinfo {author} {\bibfnamefont {F.~S.}\ \bibnamefont
  {Bergeret}}, \ and\ \bibinfo {author} {\bibfnamefont {J.~C.}\ \bibnamefont
  {Cuevas}},\ }\href@noop {} {\bibfield  {journal} {\bibinfo  {journal} {Phys.
  Rev. Lett.}\ }\textbf {\bibinfo {volume} {104}},\ \bibinfo {pages} {247003}
  (\bibinfo {year} {2010})}\BibitemShut {NoStop}%
\bibitem [{\citenamefont {Chiodi}\ \emph {et~al.}(2009)\citenamefont {Chiodi},
  \citenamefont {Aprili},\ and\ \citenamefont {Reulet}}]{ChiodiPRL2009}%
  \BibitemOpen
  \bibfield  {author} {\bibinfo {author} {\bibfnamefont {F.}~\bibnamefont
  {Chiodi}}, \bibinfo {author} {\bibfnamefont {M.}~\bibnamefont {Aprili}}, \
  and\ \bibinfo {author} {\bibfnamefont {B.}~\bibnamefont {Reulet}},\
  }\href@noop {} {\bibfield  {journal} {\bibinfo  {journal} {Phys. Rev. Lett.}\
  }\textbf {\bibinfo {volume} {103}},\ \bibinfo {pages} {177002} (\bibinfo
  {year} {2009})}\BibitemShut {NoStop}%
\bibitem [{\citenamefont {Fuechsle}\ \emph {et~al.}(2009)\citenamefont
  {Fuechsle}, \citenamefont {Bentner}, \citenamefont {Ryndyk}, \citenamefont
  {Reinwald}, \citenamefont {Wegscheider},\ and\ \citenamefont
  {Strunk}}]{FuechslePRL2009}%
  \BibitemOpen
  \bibfield  {author} {\bibinfo {author} {\bibfnamefont {M.}~\bibnamefont
  {Fuechsle}}, \bibinfo {author} {\bibfnamefont {J.}~\bibnamefont {Bentner}},
  \bibinfo {author} {\bibfnamefont {D.~A.}\ \bibnamefont {Ryndyk}}, \bibinfo
  {author} {\bibfnamefont {M.}~\bibnamefont {Reinwald}}, \bibinfo {author}
  {\bibfnamefont {W.}~\bibnamefont {Wegscheider}}, \ and\ \bibinfo {author}
  {\bibfnamefont {C.}~\bibnamefont {Strunk}},\ }\href@noop {} {\bibfield
  {journal} {\bibinfo  {journal} {Phys. Rev. Lett.}\ }\textbf {\bibinfo
  {volume} {102}},\ \bibinfo {pages} {127001} (\bibinfo {year}
  {2009})}\BibitemShut {NoStop}%
\bibitem [{\citenamefont {Chiodi}\ \emph {et~al.}(2011)\citenamefont {Chiodi},
  \citenamefont {Ferrier}, \citenamefont {Tikhonov}, \citenamefont {Virtanen},
  \citenamefont {Heikkil{\"a}}, \citenamefont {Feigelman}, \citenamefont
  {Gu\'eron},\ and\ \citenamefont {Bouchiat}}]{ChiodiSR2011}%
  \BibitemOpen
  \bibfield  {author} {\bibinfo {author} {\bibfnamefont {F.}~\bibnamefont
  {Chiodi}}, \bibinfo {author} {\bibfnamefont {M.}~\bibnamefont {Ferrier}},
  \bibinfo {author} {\bibfnamefont {K.}~\bibnamefont {Tikhonov}}, \bibinfo
  {author} {\bibfnamefont {P.}~\bibnamefont {Virtanen}}, \bibinfo {author}
  {\bibfnamefont {T.~T.}\ \bibnamefont {Heikkil{\"a}}}, \bibinfo {author}
  {\bibfnamefont {M.}~\bibnamefont {Feigelman}}, \bibinfo {author}
  {\bibfnamefont {S.}~\bibnamefont {Gu\'eron}}, \ and\ \bibinfo {author}
  {\bibfnamefont {H.}~\bibnamefont {Bouchiat}},\ }\href@noop {} {\bibfield
  {journal} {\bibinfo  {journal} {Scientific Reports}\ }\textbf {\bibinfo
  {volume} {1}},\ \bibinfo {pages} {3} (\bibinfo {year} {2011})}\BibitemShut
  {NoStop}%
\bibitem [{\citenamefont {Tikhonov}\ and\ \citenamefont
  {Feigel'man}(2015)}]{TikhonovPRB2015}%
  \BibitemOpen
  \bibfield  {author} {\bibinfo {author} {\bibfnamefont {K.~S.}\ \bibnamefont
  {Tikhonov}}\ and\ \bibinfo {author} {\bibfnamefont {M.~V.}\ \bibnamefont
  {Feigel'man}},\ }\href@noop {} {\bibfield  {journal} {\bibinfo  {journal}
  {Phys. Rev. B}\ }\textbf {\bibinfo {volume} {91}},\ \bibinfo {pages} {054519}
  (\bibinfo {year} {2015})}\BibitemShut {NoStop}%
\bibitem [{\citenamefont {Beenakker}\ and\ \citenamefont {van
  Houten}(1991)}]{BeenakkerVanHouten1991}%
  \BibitemOpen
  \bibfield  {author} {\bibinfo {author} {\bibfnamefont {C.~W.~J.}\
  \bibnamefont {Beenakker}}\ and\ \bibinfo {author} {\bibfnamefont
  {H.}~\bibnamefont {van Houten}},\ }in\ \href@noop {} {\emph {\bibinfo
  {booktitle} {Solid State Physics}}},\ Vol.~\bibinfo {volume} {44}\ (\bibinfo
  {year} {1991})\BibitemShut {NoStop}%
\bibitem [{\citenamefont {Nazarov}\ and\ \citenamefont
  {Blanter}(2009)}]{NazarovBlanter2009}%
  \BibitemOpen
  \bibfield  {author} {\bibinfo {author} {\bibfnamefont {Y.~N.}\ \bibnamefont
  {Nazarov}}\ and\ \bibinfo {author} {\bibfnamefont {Y.~M.}\ \bibnamefont
  {Blanter}},\ }\href@noop {} {\emph {\bibinfo {title} {Quantum transport.
  Introduction to Nanoscience}}}\ (\bibinfo  {publisher} {Cambridge University
  Press},\ \bibinfo {year} {2009})\BibitemShut {NoStop}%
\bibitem [{\citenamefont {Bretheau}\ \emph {et~al.}(2012)\citenamefont
  {Bretheau}, \citenamefont {Girit}, \citenamefont {Tosi}, \citenamefont
  {Goffman}, \citenamefont {Joyez}, \citenamefont {Pothier}, \citenamefont
  {Est{\`e}ve},\ and\ \citenamefont {Urbina}}]{Bretheau2012}%
  \BibitemOpen
  \bibfield  {author} {\bibinfo {author} {\bibfnamefont {L.}~\bibnamefont
  {Bretheau}}, \bibinfo {author} {\bibfnamefont {{\c C}.}~\bibnamefont
  {Girit}}, \bibinfo {author} {\bibfnamefont {L.}~\bibnamefont {Tosi}},
  \bibinfo {author} {\bibfnamefont {M.}~\bibnamefont {Goffman}}, \bibinfo
  {author} {\bibfnamefont {P.}~\bibnamefont {Joyez}}, \bibinfo {author}
  {\bibfnamefont {H.}~\bibnamefont {Pothier}}, \bibinfo {author} {\bibfnamefont
  {D.}~\bibnamefont {Est{\`e}ve}}, \ and\ \bibinfo {author} {\bibfnamefont
  {C.}~\bibnamefont {Urbina}},\ }\href@noop {} {\bibfield  {journal} {\bibinfo
  {journal} {Comptes Rendus Physique}\ }\textbf {\bibinfo {volume} {13}},\
  \bibinfo {pages} {89} (\bibinfo {year} {2012})}\BibitemShut {NoStop}%
\bibitem [{\citenamefont {Beenakker}(1991)}]{Beenakker1991}%
  \BibitemOpen
  \bibfield  {author} {\bibinfo {author} {\bibfnamefont {C.~W.~J.}\
  \bibnamefont {Beenakker}},\ }\href@noop {} {\bibfield  {journal} {\bibinfo
  {journal} {Phys. Rev. Lett.}\ }\textbf {\bibinfo {volume} {67}},\ \bibinfo
  {pages} {3836} (\bibinfo {year} {1991})}\BibitemShut {NoStop}%
\bibitem [{\citenamefont {Kulik}\ and\ \citenamefont
  {Omel'yanchuk}(1977)}]{KOII1977}%
  \BibitemOpen
  \bibfield  {author} {\bibinfo {author} {\bibfnamefont {I.~O.}\ \bibnamefont
  {Kulik}}\ and\ \bibinfo {author} {\bibfnamefont {A.~N.}\ \bibnamefont
  {Omel'yanchuk}},\ }\href@noop {} {\bibfield  {journal} {\bibinfo  {journal}
  {Soviet Journal of Low Temperature Physics}\ }\textbf {\bibinfo {volume}
  {3}},\ \bibinfo {pages} {7} (\bibinfo {year} {1977})}\BibitemShut {NoStop}%
\bibitem [{\citenamefont {Pershoguba}\ \emph {et~al.}(2019)\citenamefont
  {Pershoguba}, \citenamefont {Veness},\ and\ \citenamefont
  {Glazman}}]{Glazman2019}%
  \BibitemOpen
  \bibfield  {author} {\bibinfo {author} {\bibfnamefont {S.~S.}\ \bibnamefont
  {Pershoguba}}, \bibinfo {author} {\bibfnamefont {T.}~\bibnamefont {Veness}},
  \ and\ \bibinfo {author} {\bibfnamefont {L.~I.}\ \bibnamefont {Glazman}},\
  }\href@noop {} {\bibfield  {journal} {\bibinfo  {journal} {Phys. Rev. Lett.}\
  }\textbf {\bibinfo {volume} {123}},\ \bibinfo {pages} {067001} (\bibinfo
  {year} {2019})}\BibitemShut {NoStop}%
\bibitem [{\citenamefont {Chauvin}\ \emph {et~al.}(2007)\citenamefont
  {Chauvin}, \citenamefont {vom Stein}, \citenamefont {Esteve}, \citenamefont
  {Urbina}, \citenamefont {Cuevas},\ and\ \citenamefont
  {Levy~Yeyati}}]{Chauvin2007}%
  \BibitemOpen
  \bibfield  {author} {\bibinfo {author} {\bibfnamefont {M.}~\bibnamefont
  {Chauvin}}, \bibinfo {author} {\bibfnamefont {P.}~\bibnamefont {vom Stein}},
  \bibinfo {author} {\bibfnamefont {D.}~\bibnamefont {Esteve}}, \bibinfo
  {author} {\bibfnamefont {C.}~\bibnamefont {Urbina}}, \bibinfo {author}
  {\bibfnamefont {J.~C.}\ \bibnamefont {Cuevas}}, \ and\ \bibinfo {author}
  {\bibfnamefont {A.}~\bibnamefont {Levy~Yeyati}},\ }\href@noop {} {\bibfield
  {journal} {\bibinfo  {journal} {Phys. Rev. Lett.}\ }\textbf {\bibinfo
  {volume} {99}},\ \bibinfo {pages} {067008} (\bibinfo {year}
  {2007})}\BibitemShut {NoStop}%
\bibitem [{\citenamefont {Della~Rocca}\ \emph {et~al.}(2007)\citenamefont
  {Della~Rocca}, \citenamefont {Chauvin}, \citenamefont {Huard}, \citenamefont
  {Pothier}, \citenamefont {Esteve},\ and\ \citenamefont
  {Urbina}}]{DellaRocca2007}%
  \BibitemOpen
  \bibfield  {author} {\bibinfo {author} {\bibfnamefont {M.~L.}\ \bibnamefont
  {Della~Rocca}}, \bibinfo {author} {\bibfnamefont {M.}~\bibnamefont
  {Chauvin}}, \bibinfo {author} {\bibfnamefont {B.}~\bibnamefont {Huard}},
  \bibinfo {author} {\bibfnamefont {H.}~\bibnamefont {Pothier}}, \bibinfo
  {author} {\bibfnamefont {D.}~\bibnamefont {Esteve}}, \ and\ \bibinfo {author}
  {\bibfnamefont {C.}~\bibnamefont {Urbina}},\ }\href@noop {} {\bibfield
  {journal} {\bibinfo  {journal} {Phys. Rev. Lett.}\ }\textbf {\bibinfo
  {volume} {99}},\ \bibinfo {pages} {127005} (\bibinfo {year}
  {2007})}\BibitemShut {NoStop}%
\bibitem [{\citenamefont {Bretheau}\ \emph {et~al.}(2013)\citenamefont
  {Bretheau}, \citenamefont {Girit}, \citenamefont {Urbina}, \citenamefont
  {Esteve},\ and\ \citenamefont {Pothier}}]{Bretheau2013}%
  \BibitemOpen
  \bibfield  {author} {\bibinfo {author} {\bibfnamefont {L.}~\bibnamefont
  {Bretheau}}, \bibinfo {author} {\bibfnamefont {{\c C}.~{\"O}.}\ \bibnamefont
  {Girit}}, \bibinfo {author} {\bibfnamefont {C.}~\bibnamefont {Urbina}},
  \bibinfo {author} {\bibfnamefont {D.}~\bibnamefont {Esteve}}, \ and\ \bibinfo
  {author} {\bibfnamefont {H.}~\bibnamefont {Pothier}},\ }\href@noop {}
  {\bibfield  {journal} {\bibinfo  {journal} {Physical Review X}\ }\textbf
  {\bibinfo {volume} {3}},\ \bibinfo {pages} {041034} (\bibinfo {year}
  {2013})}\BibitemShut {NoStop}%
\bibitem [{\citenamefont {Wiedenmann}\ \emph {et~al.}(2016)\citenamefont
  {Wiedenmann}, \citenamefont {Bocquillon}, \citenamefont {Deacon},
  \citenamefont {Hartinger}, \citenamefont {Hermann}, \citenamefont {Klapwijk},
  \citenamefont {Maier}, \citenamefont {Ames}, \citenamefont {Br\"une},
  \citenamefont {Gould}, \citenamefont {Oiwa}, \citenamefont {Tarucha},
  \citenamefont {Buhmann},\ and\ \citenamefont {Molenkamp}}]{Wiedenmann2016}%
  \BibitemOpen
  \bibfield  {author} {\bibinfo {author} {\bibfnamefont {J.}~\bibnamefont
  {Wiedenmann}}, \bibinfo {author} {\bibfnamefont {E.}~\bibnamefont
  {Bocquillon}}, \bibinfo {author} {\bibfnamefont {R.~S.}\ \bibnamefont
  {Deacon}}, \bibinfo {author} {\bibfnamefont {S.}~\bibnamefont {Hartinger}},
  \bibinfo {author} {\bibfnamefont {O.}~\bibnamefont {Hermann}}, \bibinfo
  {author} {\bibfnamefont {T.~M.}\ \bibnamefont {Klapwijk}}, \bibinfo {author}
  {\bibfnamefont {L.}~\bibnamefont {Maier}}, \bibinfo {author} {\bibfnamefont
  {C.}~\bibnamefont {Ames}}, \bibinfo {author} {\bibfnamefont {C.}~\bibnamefont
  {Br\"une}}, \bibinfo {author} {\bibfnamefont {C.}~\bibnamefont {Gould}},
  \bibinfo {author} {\bibfnamefont {A.}~\bibnamefont {Oiwa}}, \bibinfo {author}
  {\bibfnamefont {S.}~\bibnamefont {Tarucha}}, \bibinfo {author} {\bibfnamefont
  {H.}~\bibnamefont {Buhmann}}, \ and\ \bibinfo {author} {\bibfnamefont
  {L.~W.}\ \bibnamefont {Molenkamp}},\ }\href@noop {} {\bibfield  {journal}
  {\bibinfo  {journal} {Nature Communications}\ ,\ \bibinfo {pages} {10303}}
  (\bibinfo {year} {2016})}\BibitemShut {NoStop}%
\bibitem [{\citenamefont {Bocquillon}\ \emph {et~al.}(2016)\citenamefont
  {Bocquillon}, \citenamefont {Deacon}, \citenamefont {Wiedenmann},
  \citenamefont {Leubner}, \citenamefont {Klapwijk}, \citenamefont {Br\"une},
  \citenamefont {Ishibashi}, \citenamefont {Buhmann},\ and\ \citenamefont
  {Molenkamp}}]{Bocquillon2016}%
  \BibitemOpen
  \bibfield  {author} {\bibinfo {author} {\bibfnamefont {E.}~\bibnamefont
  {Bocquillon}}, \bibinfo {author} {\bibfnamefont {R.~S.}\ \bibnamefont
  {Deacon}}, \bibinfo {author} {\bibfnamefont {J.}~\bibnamefont {Wiedenmann}},
  \bibinfo {author} {\bibfnamefont {P.}~\bibnamefont {Leubner}}, \bibinfo
  {author} {\bibfnamefont {T.~M.}\ \bibnamefont {Klapwijk}}, \bibinfo {author}
  {\bibfnamefont {C.}~\bibnamefont {Br\"une}}, \bibinfo {author} {\bibfnamefont
  {K.}~\bibnamefont {Ishibashi}}, \bibinfo {author} {\bibfnamefont
  {H.}~\bibnamefont {Buhmann}}, \ and\ \bibinfo {author} {\bibfnamefont
  {L.~W.}\ \bibnamefont {Molenkamp}},\ }\href@noop {} {\bibfield  {journal}
  {\bibinfo  {journal} {Nature Nanotechnology}\ }\textbf {\bibinfo {volume}
  {12}},\ \bibinfo {pages} {159} (\bibinfo {year} {2016})}\BibitemShut
  {NoStop}%
\bibitem [{\citenamefont {Deacon}\ \emph {et~al.}(2017)\citenamefont {Deacon},
  \citenamefont {Wiedenmann}, \citenamefont {Bocquillon}, \citenamefont
  {Dom\'inguez}, \citenamefont {Klapwijk}, \citenamefont {Leubner},
  \citenamefont {Br\"une}, \citenamefont {Hankiewicz}, \citenamefont {Tarucha},
  \citenamefont {Ishibashi}, \citenamefont {Buhmann},\ and\ \citenamefont
  {Molenkamp}}]{Deacon2017}%
  \BibitemOpen
  \bibfield  {author} {\bibinfo {author} {\bibfnamefont {R.~S.}\ \bibnamefont
  {Deacon}}, \bibinfo {author} {\bibfnamefont {J.}~\bibnamefont {Wiedenmann}},
  \bibinfo {author} {\bibfnamefont {E.}~\bibnamefont {Bocquillon}}, \bibinfo
  {author} {\bibfnamefont {F.}~\bibnamefont {Dom\'inguez}}, \bibinfo {author}
  {\bibfnamefont {T.~M.}\ \bibnamefont {Klapwijk}}, \bibinfo {author}
  {\bibfnamefont {P.}~\bibnamefont {Leubner}}, \bibinfo {author} {\bibfnamefont
  {C.}~\bibnamefont {Br\"une}}, \bibinfo {author} {\bibfnamefont {E.~M.}\
  \bibnamefont {Hankiewicz}}, \bibinfo {author} {\bibfnamefont
  {S.}~\bibnamefont {Tarucha}}, \bibinfo {author} {\bibfnamefont
  {K.}~\bibnamefont {Ishibashi}}, \bibinfo {author} {\bibfnamefont
  {H.}~\bibnamefont {Buhmann}}, \ and\ \bibinfo {author} {\bibfnamefont
  {L.~W.}\ \bibnamefont {Molenkamp}},\ }\href@noop {} {\bibfield  {journal}
  {\bibinfo  {journal} {Physical Review X}\ }\textbf {\bibinfo {volume} {7}},\
  \bibinfo {pages} {021011} (\bibinfo {year} {2017})}\BibitemShut {NoStop}%
\bibitem [{\citenamefont {Dom\'inguez}\ \emph {et~al.}(2017)\citenamefont
  {Dom\'inguez}, \citenamefont {Kashuba}, \citenamefont {Bocquillon},
  \citenamefont {Wiedenmann}, \citenamefont {Deacon}, \citenamefont {Klapwijk},
  \citenamefont {Platero}, \citenamefont {Molenkamp}, \citenamefont
  {Trauzettel},\ and\ \citenamefont {Hankiewicz}}]{Dominguez2017}%
  \BibitemOpen
  \bibfield  {author} {\bibinfo {author} {\bibfnamefont {F.}~\bibnamefont
  {Dom\'inguez}}, \bibinfo {author} {\bibfnamefont {O.}~\bibnamefont
  {Kashuba}}, \bibinfo {author} {\bibfnamefont {E.}~\bibnamefont {Bocquillon}},
  \bibinfo {author} {\bibfnamefont {J.}~\bibnamefont {Wiedenmann}}, \bibinfo
  {author} {\bibfnamefont {R.~S.}\ \bibnamefont {Deacon}}, \bibinfo {author}
  {\bibfnamefont {T.~M.}\ \bibnamefont {Klapwijk}}, \bibinfo {author}
  {\bibfnamefont {G.}~\bibnamefont {Platero}}, \bibinfo {author} {\bibfnamefont
  {L.~W.}\ \bibnamefont {Molenkamp}}, \bibinfo {author} {\bibfnamefont
  {B.}~\bibnamefont {Trauzettel}}, \ and\ \bibinfo {author} {\bibfnamefont
  {E.~M.}\ \bibnamefont {Hankiewicz}},\ }\href@noop {} {\bibfield  {journal}
  {\bibinfo  {journal} {Phys. Rev. B}\ }\textbf {\bibinfo {volume} {95}},\
  \bibinfo {pages} {195430} (\bibinfo {year} {2017})}\BibitemShut {NoStop}%
\bibitem [{\citenamefont {Pic{\'o}-Cort{\'e}s}\ \emph
  {et~al.}(2017)\citenamefont {Pic{\'o}-Cort{\'e}s}, \citenamefont
  {Dom\'inguez},\ and\ \citenamefont {Platero}}]{PicoCortesPRB2017}%
  \BibitemOpen
  \bibfield  {author} {\bibinfo {author} {\bibfnamefont {J.}~\bibnamefont
  {Pic{\'o}-Cort{\'e}s}}, \bibinfo {author} {\bibfnamefont {F.}~\bibnamefont
  {Dom\'inguez}}, \ and\ \bibinfo {author} {\bibfnamefont {G.}~\bibnamefont
  {Platero}},\ }\href@noop {} {\bibfield  {journal} {\bibinfo  {journal} {Phys.
  Rev. B}\ }\textbf {\bibinfo {volume} {96}},\ \bibinfo {pages} {125438}
  (\bibinfo {year} {2017})}\BibitemShut {NoStop}%
\bibitem [{\citenamefont {Bergeret}\ \emph {et~al.}(2010)\citenamefont
  {Bergeret}, \citenamefont {Virtanen}, \citenamefont {Heikkil{\"a}},\ and\
  \citenamefont {Cuevas}}]{Bergeret2010}%
  \BibitemOpen
  \bibfield  {author} {\bibinfo {author} {\bibfnamefont {F.~S.}\ \bibnamefont
  {Bergeret}}, \bibinfo {author} {\bibfnamefont {P.}~\bibnamefont {Virtanen}},
  \bibinfo {author} {\bibfnamefont {T.~T.}\ \bibnamefont {Heikkil{\"a}}}, \
  and\ \bibinfo {author} {\bibfnamefont {J.~C.}\ \bibnamefont {Cuevas}},\
  }\href@noop {} {\bibfield  {journal} {\bibinfo  {journal} {Phys. Rev. Lett.}\
  }\textbf {\bibinfo {volume} {105}},\ \bibinfo {pages} {117001} (\bibinfo
  {year} {2010})}\BibitemShut {NoStop}%
\bibitem [{\citenamefont {Semenov}\ \emph {et~al.}(2016)\citenamefont
  {Semenov}, \citenamefont {Devyatov}, \citenamefont {de~Visser},\ and\
  \citenamefont {Klapwijk}}]{Semenov2016}%
  \BibitemOpen
  \bibfield  {author} {\bibinfo {author} {\bibfnamefont {A.~V.}\ \bibnamefont
  {Semenov}}, \bibinfo {author} {\bibfnamefont {I.~A.}\ \bibnamefont
  {Devyatov}}, \bibinfo {author} {\bibfnamefont {P.~J.}\ \bibnamefont
  {de~Visser}}, \ and\ \bibinfo {author} {\bibfnamefont {T.~M.}\ \bibnamefont
  {Klapwijk}},\ }\href@noop {} {\bibfield  {journal} {\bibinfo  {journal}
  {Phys. Rev. Lett.}\ }\textbf {\bibinfo {volume} {117}},\ \bibinfo {pages}
  {047002} (\bibinfo {year} {2016})}\BibitemShut {NoStop}%
\bibitem [{\citenamefont {Semenov}\ \emph {et~al.}(2020)\citenamefont
  {Semenov}, \citenamefont {Devyatov}, \citenamefont {Westig},\ and\
  \citenamefont {Klapwijk}}]{Semenov2019}%
  \BibitemOpen
  \bibfield  {author} {\bibinfo {author} {\bibfnamefont {A.~V.}\ \bibnamefont
  {Semenov}}, \bibinfo {author} {\bibfnamefont {I.~A.}\ \bibnamefont
  {Devyatov}}, \bibinfo {author} {\bibfnamefont {M.~P.}\ \bibnamefont
  {Westig}}, \ and\ \bibinfo {author} {\bibfnamefont {T.~M.}\ \bibnamefont
  {Klapwijk}},\ }\href@noop {} {\bibfield  {journal} {\bibinfo  {journal}
  {Physical Review Applied, in press}\ } (\bibinfo {year} {2020})}\BibitemShut
  {NoStop}%
\bibitem [{\citenamefont {Pals}\ and\ \citenamefont
  {Dobben}(1980)}]{PalsDobben1980}%
  \BibitemOpen
  \bibfield  {author} {\bibinfo {author} {\bibfnamefont {J.~A.}\ \bibnamefont
  {Pals}}\ and\ \bibinfo {author} {\bibfnamefont {J.}~\bibnamefont {Dobben}},\
  }\href@noop {} {\bibfield  {journal} {\bibinfo  {journal} {Phys. Rev. Lett.}\
  }\textbf {\bibinfo {volume} {44}},\ \bibinfo {pages} {1143} (\bibinfo {year}
  {1980})}\BibitemShut {NoStop}%
\bibitem [{\citenamefont {Horstman}\ and\ \citenamefont
  {Wolter}(1981)}]{HorstmanWolter1981}%
  \BibitemOpen
  \bibfield  {author} {\bibinfo {author} {\bibfnamefont {R.~E.}\ \bibnamefont
  {Horstman}}\ and\ \bibinfo {author} {\bibfnamefont {J.}~\bibnamefont
  {Wolter}},\ }\href@noop {} {\bibfield  {journal} {\bibinfo  {journal}
  {Physics Letters A}\ }\textbf {\bibinfo {volume} {82A}},\ \bibinfo {pages}
  {43} (\bibinfo {year} {1981})}\BibitemShut {NoStop}%
\bibitem [{\citenamefont {Wolter}\ and\ \citenamefont
  {Horstman}(1981)}]{WolterHorstman1981}%
  \BibitemOpen
  \bibfield  {author} {\bibinfo {author} {\bibfnamefont {J.}~\bibnamefont
  {Wolter}}\ and\ \bibinfo {author} {\bibfnamefont {R.~E.}\ \bibnamefont
  {Horstman}},\ }\href@noop {} {\bibfield  {journal} {\bibinfo  {journal}
  {Physics Letters A}\ }\textbf {\bibinfo {volume} {86A}},\ \bibinfo {pages}
  {185} (\bibinfo {year} {1981})}\BibitemShut {NoStop}%
\bibitem [{\citenamefont {Cavalleri}(2018)}]{Cavalleri2017}%
  \BibitemOpen
  \bibfield  {author} {\bibinfo {author} {\bibfnamefont {A.}~\bibnamefont
  {Cavalleri}},\ }\href {\doibase 10.1080/00107514.2017.1406623} {\bibfield
  {journal} {\bibinfo  {journal} {Contemporary Physics}\ }\textbf {\bibinfo
  {volume} {59}},\ \bibinfo {pages} {31} (\bibinfo {year} {2018})},\ \Eprint
  {http://arxiv.org/abs/https://doi.org/10.1080/00107514.2017.1406623}
  {https://doi.org/10.1080/00107514.2017.1406623} \BibitemShut {NoStop}%
\bibitem [{\citenamefont {Fausti}\ \emph {et~al.}(2011)\citenamefont {Fausti},
  \citenamefont {Tobey}, \citenamefont {Dean}, \citenamefont {Kaiser},
  \citenamefont {Dienst}, \citenamefont {Hoffmann}, \citenamefont {Pyon},
  \citenamefont {Takayama}, \citenamefont {Takagi},\ and\ \citenamefont
  {Cavalleri}}]{Fausti2011}%
  \BibitemOpen
  \bibfield  {author} {\bibinfo {author} {\bibfnamefont {D.}~\bibnamefont
  {Fausti}}, \bibinfo {author} {\bibfnamefont {R.~I.}\ \bibnamefont {Tobey}},
  \bibinfo {author} {\bibfnamefont {N.}~\bibnamefont {Dean}}, \bibinfo {author}
  {\bibfnamefont {S.}~\bibnamefont {Kaiser}}, \bibinfo {author} {\bibfnamefont
  {A.}~\bibnamefont {Dienst}}, \bibinfo {author} {\bibfnamefont {M.~C.}\
  \bibnamefont {Hoffmann}}, \bibinfo {author} {\bibfnamefont {S.}~\bibnamefont
  {Pyon}}, \bibinfo {author} {\bibfnamefont {T.}~\bibnamefont {Takayama}},
  \bibinfo {author} {\bibfnamefont {H.}~\bibnamefont {Takagi}}, \ and\ \bibinfo
  {author} {\bibfnamefont {A.}~\bibnamefont {Cavalleri}},\ }\href@noop {}
  {\bibfield  {journal} {\bibinfo  {journal} {Science}\ }\textbf {\bibinfo
  {volume} {331}},\ \bibinfo {pages} {189} (\bibinfo {year}
  {2011})}\BibitemShut {NoStop}%
\bibitem [{\citenamefont {Mitrano}\ \emph {et~al.}(2016)\citenamefont
  {Mitrano}, \citenamefont {Cantaluppi}, \citenamefont {Nicoletti},
  \citenamefont {Kaiser}, \citenamefont {Perucchi}, \citenamefont {Lupi},
  \citenamefont {Di~Pietro}, \citenamefont {Pontiroli}, \citenamefont
  {Ricc{\`o}}, \citenamefont {Clark}, \citenamefont {Jaksch},\ and\
  \citenamefont {Cavalleri}}]{MitranoNature2016}%
  \BibitemOpen
  \bibfield  {author} {\bibinfo {author} {\bibfnamefont {M.}~\bibnamefont
  {Mitrano}}, \bibinfo {author} {\bibfnamefont {A.}~\bibnamefont {Cantaluppi}},
  \bibinfo {author} {\bibfnamefont {D.}~\bibnamefont {Nicoletti}}, \bibinfo
  {author} {\bibfnamefont {S.}~\bibnamefont {Kaiser}}, \bibinfo {author}
  {\bibfnamefont {A.}~\bibnamefont {Perucchi}}, \bibinfo {author}
  {\bibfnamefont {S.}~\bibnamefont {Lupi}}, \bibinfo {author} {\bibfnamefont
  {P.}~\bibnamefont {Di~Pietro}}, \bibinfo {author} {\bibfnamefont
  {D.}~\bibnamefont {Pontiroli}}, \bibinfo {author} {\bibfnamefont
  {M.}~\bibnamefont {Ricc{\`o}}}, \bibinfo {author} {\bibfnamefont {S.~R.}\
  \bibnamefont {Clark}}, \bibinfo {author} {\bibfnamefont {D.}~\bibnamefont
  {Jaksch}}, \ and\ \bibinfo {author} {\bibfnamefont {A.}~\bibnamefont
  {Cavalleri}},\ }\href@noop {} {\bibfield  {journal} {\bibinfo  {journal}
  {Nature}\ }\textbf {\bibinfo {volume} {530}},\ \bibinfo {pages} {461}
  (\bibinfo {year} {2016})}\BibitemShut {NoStop}%
\bibitem [{\citenamefont {Demsar}(2016)}]{Demsar2016}%
  \BibitemOpen
  \bibfield  {author} {\bibinfo {author} {\bibfnamefont {J.}~\bibnamefont
  {Demsar}},\ }\href@noop {} {\bibfield  {journal} {\bibinfo  {journal} {Nature
  Physics}\ }\textbf {\bibinfo {volume} {12}},\ \bibinfo {pages} {202}
  (\bibinfo {year} {2016})}\BibitemShut {NoStop}%
\bibitem [{\citenamefont {Beck}\ \emph {et~al.}(2013)\citenamefont {Beck},
  \citenamefont {Rousseau}, \citenamefont {Klammer}, \citenamefont {Leiderer},
  \citenamefont {Mittendorff}, \citenamefont {Winnerl}, \citenamefont {Helm},
  \citenamefont {Gol'tsman},\ and\ \citenamefont {Demsar}}]{BeckPRL2013}%
  \BibitemOpen
  \bibfield  {author} {\bibinfo {author} {\bibfnamefont {M.}~\bibnamefont
  {Beck}}, \bibinfo {author} {\bibfnamefont {I.}~\bibnamefont {Rousseau}},
  \bibinfo {author} {\bibfnamefont {M.}~\bibnamefont {Klammer}}, \bibinfo
  {author} {\bibfnamefont {P.}~\bibnamefont {Leiderer}}, \bibinfo {author}
  {\bibfnamefont {M.}~\bibnamefont {Mittendorff}}, \bibinfo {author}
  {\bibfnamefont {S.}~\bibnamefont {Winnerl}}, \bibinfo {author} {\bibfnamefont
  {M.}~\bibnamefont {Helm}}, \bibinfo {author} {\bibfnamefont {G.~N.}\
  \bibnamefont {Gol'tsman}}, \ and\ \bibinfo {author} {\bibfnamefont
  {J.}~\bibnamefont {Demsar}},\ }\href@noop {} {\bibfield  {journal} {\bibinfo
  {journal} {Phys. Rev. Lett.}\ }\textbf {\bibinfo {volume} {110}},\ \bibinfo
  {pages} {267003} (\bibinfo {year} {2013})}\BibitemShut {NoStop}%
\bibitem [{\citenamefont {Han}\ \emph {et~al.}(2014)\citenamefont {Han},
  \citenamefont {Allain}, \citenamefont {Arjmandi-Tash}, \citenamefont
  {Tikhonov}, \citenamefont {Feigel'man}, \citenamefont {Sac{\'e}p{\'e}},\ and\
  \citenamefont {Bouchiat}}]{Han14}%
  \BibitemOpen
  \bibfield  {author} {\bibinfo {author} {\bibfnamefont {Z.}~\bibnamefont
  {Han}}, \bibinfo {author} {\bibfnamefont {A.}~\bibnamefont {Allain}},
  \bibinfo {author} {\bibfnamefont {H.}~\bibnamefont {Arjmandi-Tash}}, \bibinfo
  {author} {\bibfnamefont {K.}~\bibnamefont {Tikhonov}}, \bibinfo {author}
  {\bibfnamefont {M.}~\bibnamefont {Feigel'man}}, \bibinfo {author}
  {\bibfnamefont {B.}~\bibnamefont {Sac{\'e}p{\'e}}}, \ and\ \bibinfo {author}
  {\bibfnamefont {V.}~\bibnamefont {Bouchiat}},\ }\href@noop {} {\bibfield
  {journal} {\bibinfo  {journal} {Nature Physics}\ }\textbf {\bibinfo {volume}
  {10}},\ \bibinfo {pages} {380} (\bibinfo {year} {2014})}\BibitemShut
  {NoStop}%
\bibitem [{\citenamefont {Sac{\'e}p{\'e}}\ \emph {et~al.}(2020)\citenamefont
  {Sac{\'e}p{\'e}}, \citenamefont {Feigel'man},\ and\ \citenamefont
  {Klapwijk}}]{SacepeReview2020}%
  \BibitemOpen
  \bibfield  {author} {\bibinfo {author} {\bibfnamefont {B.}~\bibnamefont
  {Sac{\'e}p{\'e}}}, \bibinfo {author} {\bibfnamefont {M.~V.}\ \bibnamefont
  {Feigel'man}}, \ and\ \bibinfo {author} {\bibfnamefont {T.~M.}\ \bibnamefont
  {Klapwijk}},\ }\href@noop {} {\enquote {\bibinfo {title} {Quantum breakdown
  of superconductivity in low-dimensional materials},}\ } (\bibinfo {year}
  {2020}),\ \bibinfo {note} {review in preparation}\BibitemShut {NoStop}%
\end{thebibliography}
\end{document}